\documentclass[useAMS,usenatbib,usegraphicx,fleqn]{mn2e}

\usepackage{epsfig}
\usepackage{amssymb}
\usepackage{myaasmacros}
\title[Relaxation and Stripping]{Relaxation and Stripping: The
  Evolution of Sizes, Dispersions and Dark Matter Fractions in Major and Minor
  Mergers of Elliptical Galaxies} 
\author[Hilz et al.]{Michael Hilz$^{1,2}$,Thorsten Naab$^{1}$\thanks{E-mail:
naab@mpa-garching.mpg.de},Jeremiah P. Ostriker$^{3}$,Jens Thomas$^{4}$,
\newauthor  Andreas Burkert$^{2}$ and Roland Jesseit$^{5}$\\
$^{1}$Max-Planck-Institut f\"ur Astrophysik, Karl-Schwarzschild-Str. 1, 85741 Garching, Germany \\
$^{2}$Universit\"ats-Sternwarte M\"unchen, Scheinerstr. 1 ,D-81679 M\"unchen, Germany \\
$^{3}$Department of Astrophysical Sciences, Princeton University, Princeton, NJ 08544, USA\\
$^{4}$Max-Planck-Institut f\"ur extraterrestrische Physik, Giessenbachstrasse, D-85748 Garching, Germany \\
$^{5}$Lifeware SA, Asylstrasse 112,CH-8032 Z\"urich, Switzerland}

\begin{document}

\date{Accepted ???. Received ??? in original form ???}

\pagerange{\pageref{firstpage}--\pageref{lastpage}} \pubyear{2008}

\maketitle

\label{firstpage}

\begin{abstract}
We revisit collisionless major and minor mergers of spheroidal
galaxies in the context of recent observational insights on compact
massive early-type galaxies at high redshift and 
their rapid evolution on cosmological timescales. The simulations are
performed as a series of mergers with mass-ratios of 1:1 and 1:10 for
models representing pure bulges as well as bulges embedded in dark
matter halos. For major and minor mergers, respectively, we identify
and analyze two different processes, violent relaxation and stripping,
leading to size evolution and a change of the dark matter fraction
within the observable effective radius $r_\mathrm{e}$. Violent
relaxation - which is the dominant mixing process for major mergers
but less important for minor  mergers - scatters relatively more dark
matter particles than bulge particles to radii $r <
r_\mathrm{e}$. Stripping in minor mergers assembles stellar satellite
particles at large radii in halo dominated regions of the massive
host. This strongly increases the size of the bulge into regions with
higher dark matter fractions leaving the inner host structure
almost unchanged. A factor of two mass increase by minor mergers
increases the dark matter fraction within the effective radius by 80
per cent whereas relaxation in one equal-mass merger only leads to an
increase of 25 percent. We present
analytic corrections to simple one-component virial estimates for the
evolution of the gravitational radii. These estimates are shown to
underpredict the evolution of the effective radii for parabolic minor
mergers of bulges embedded in massive dark matter halos. If such 
a two-component system grows by minor mergers alone its size growth,
$r_{\mathrm{e}} \propto M^\alpha$, reaches values of $\alpha \approx 2.4$,
significantly exceeding the simple theoretical limit of $\alpha =
2$. For major mergers the sizes grow with $\alpha \lesssim 1$. In addition, we
discuss the velocity dispersion evolution and velocity anisotropy
profiles. Our results indicate that minor mergers of galaxies embedded
in massive dark matter halos provide a potential mechanism for explaining
the rapid size growth and the build-up of massive elliptical systems
predicting significant dark matter fractions and radially biased
velocity dispersions at large radii.  
\end{abstract}


\section{INTRODUCTION}
\label{intro}

Recent observations have revealed a population of very compact,
massive  ($\approx 10^{11} M_{\odot}$) and quiescent galaxies at
z$\sim$2 with sizes of about $\approx 1 \rm kpc$
\citep{2005ApJ...626..680D,2006ApJ...650...18T,2007MNRAS.374..614L,2007ApJ...671..285T,2007ApJ...656...66Z,2007MNRAS.382..109T,2007ApJ...656...66Z,2008ApJ...687L..61B,2008ApJ...677L...5V,2008A&A...482...21C,2008ApJ...688..770F, 
2009MNRAS.392..718S,2009ApJ...695..101D,2009ApJ...697.1290B,2010ApJ...717L.103N,2012ApJ...749..121S,2012arXiv1205.4058M}. 
Elliptical galaxies of a similar mass today are larger by a factor of
4 - 5 \citep{2008ApJ...688...48V} with at least an order of magnitude
lower effective densities and significantly lower velocity dispersions
than their high-redshift counterparts
\citep{2005ApJ...631..145V,2008ApJ...688...48V,2009ApJ...704L..34C,2009ApJ...696L..43C,2009Natur.460..717V,2011arXiv1104.3860V,2011arXiv1111.3361S,2011ApJ...738L..22M}.
The measured small effective radii are most likely not caused by
observational limitations, although the low density material in the
outer parts of distant galaxies is difficult to detect
(\citealt{2009MNRAS.398..898H}). Their clustering, number densities
and core properties indicate that they are probably the progenitors
of the most massive ellipticals and Brightest Cluster Galaxies today
\citep{2009MNRAS.398..898H,2009ApJ...697.1290B}. Such compact massive galaxies
are extremely rare in the nearby universe
(\citealp{2009ApJ...692L.118T,2010ApJ...720..723T}, see also
\citealp{2010ApJ...721L..19V,2012ApJ...745..130R,2012arXiv1203.1317J,2012MNRAS.tmp.2790F,2012ApJ...751...45T}).  

The high phase-space densities of these compact $z = 2-3$ systems
imply a formation mechanism that is dominated by dissipational processes 
\citep{2005MNRAS.363....2K,2007ApJ...658..710N,2009ApJ...699L.178N,2009ApJ...692L...1J,2009ApJ...703..785D,2009MNRAS.395..160K,2009ApJ...697L..38J,2010ApJ...725.2312O,2010ApJ...722.1666W,2011ApJ...730....4B,2012ApJ...744...63O,2012arXiv1202.3441J}. 
Frequent stellar minor and major mergers, as also expected in a
cosmological context, are promising physical mechanism to explain the
subsequent rapid size growth in the absence of significant additional
dissipation and star formation  
\citep{2000MNRAS.319..168C,2006ApJ...648L..21K,2006MNRAS.366..499D,  
2008MNRAS.384....2G,2009ApJ...697.1290B,2010MNRAS.401.1099H,2011MNRAS.415.3903T,2012MNRAS.419.3018C,2012arXiv1205.4058M}. 

Observations and theoretical studies provide evidence that massive
early-type galaxies undergo on average about one major merger since
redshift $\sim$ 2 but significantly more minor mergers per unit time 
\citep{2006ApJ...652..270B,2006ApJ...648L..21K,2006ApJ...640..241B,2008ApJ...688..789G,2011ApJ...742..103L}. 
However, merger rates are notoriously difficult to measure and the
uncertainties are still large, but assuming a canonical value of one
major merger since $z = 2$, this would not be sufficient to explain the
observed size evolution
\citep{2009ApJ...697.1290B,2011arXiv1102.3398T,2011ApJ...738L..25W}. In  
addition, major mergers are highly stochastic and some massive
galaxies should have experienced no major merger at all, and would
therefore still be compact today. Direct numerical simulations in a
full cosmological context support the importance of numerous minor
mergers for the assembly of massive galaxies whose dissipative
formation phase is followed by a second phase dominated by stellar
accretion (predominantly minor mergers) onto the galaxy, driving the
size evolution
\citep{2006ApJ...648L..21K,2009ApJ...699L.178N,2010ApJ...725.2312O,2012MNRAS.tmp.2423L}. 
This theoretical finding is supported by recent direct observational and
circumstantial evidence that has recently been presented in support of 
minor mergers \citep{2010ApJ...709.1018V,2011arXiv1102.3398T,2011ApJ...738L..25W,2012MNRAS.419.3018C}.

Minor mergers are particularly efficient in reducing the effective stellar densities,
mildly reducing the velocity dispersions, and rapidly 
increasing the sizes, building up extended stellar envelopes  
(\citealp{1980ApJ...235..421M,1983MNRAS.204..219V,1983ApJ...265..597F,2009ApJ...699L.178N,2007A&A...476.1179B,2009ApJ...697.1290B,2010MNRAS.401.1099H,2010ApJ...725.2312O,2012ApJ...744...63O,2012arXiv1202.2357L,2010ApJ...712...88L}). However,
there are some doubts as to whether minor mergers alone are sufficient
or whether additional physical processes are required
\citep{2006ApJ...636L..81N,2009ApJ...706L..86N,2012arXiv1202.0971N,2012ApJ...746..162N,2012arXiv1202.5403C}. Major
mergers of ellipticals will contribute to mass growth and will change
kinematic properties \citep{2006ApJ...636L..81N}. However, their
impact on the evolution of stellar densities, velocity dispersions and
sizes is relatively weak
\citep{1978MNRAS.184..185W,1979MNRAS.189..831W,2005MNRAS.362..184B,2009ApJ...706L..86N}.

Using the virial theorem, \citet{2000MNRAS.319..168C},
\citet{2009ApJ...699L.178N} 
and \citet{2009ApJ...697.1290B} presented a simple analytical estimate
of how sizes, densities and velocity dispersions of one-component
collisionless systems evolve during parabolic, purely stellar mergers
with different mass ratios. According to this simplified model, the
accretion of loosely bound material (minor mergers) results in a
significantly stronger size increase than predicted for major mergers
\citep{2009ApJ...699L.178N}. Using a similar approach
\citet{2009ApJ...697.1290B} argued that eight successive mergers of
mass ratio 1:10 can lead to a size increase of $\sim$ 5,  which
corresponds to the observed difference between old compact galaxies
and massive ellipticals today. Of course, this is only valid for
global system properties like the gravitational radii and total mean
square speeds. The simple virial estimates presented by
\citet{2009ApJ...699L.178N,2009ApJ...697.1290B} did not include
violent relaxation effects like mass loss, occurring during the
encounter or non-homology. However, \citet{2007ApJ...658...65C}
presented a more general study including the effect of gas dissipation
and non-homology.   

Early papers on the interactions of spheroidal galaxies already
discussed many of the aforementioned effects using N-body simulations
of one-component spherical systems.
\citet{1978MNRAS.184..185W,1979MNRAS.189..831W}, who made one of the
first simulations of this kind, already found that relaxation effects
are important in equal-mass encounters and change the internal
structure of the remnants. The central regions contract and diffuse
envelopes build up  (see also
\citealt{1980ApJ...235..421M,1983MNRAS.204..219V,1983ApJ...265..597F}),
leading to a break in homology. Furthermore, equal-mass mergers
lead to the redistribution of particles eventually reducing population
gradients \citep{1980MNRAS.191P...1W,1983MNRAS.204..219V}. This is different 
in unequal-mass mergers, which can enhance population gradients by
depositing satellite stars at large radii
\citep{1983MNRAS.204..219V}. 
%\citet{1983ApJ...265..597F} also showed,that their multiple
%equal-mass mergers nicely recover the Faber-Jackson relation
%\citep{1976ApJ...204..668F} and that the velocity dispersion becomes
%radially biased in the outer regions of the newly developed extended
%envelope. 
However, these early models did not investigate the influence of an
extended and massive dark matter halo.

%\citet{2006MNRAS.369..625N} and \citet{2009ApJ...691.1424H} later on
%showed, that more realistic galaxy models, where the bulge is
%embedded in a dark matter halo, can change the size increase. 

Although dissipationless minor mergers theoretically lead to an
increase in sizes and a decrease of velocity dispersions, it is not
clear whether this scenario works
quantitatively. \citet{2003MNRAS.342..501N}, using two-component
models - argued that dry major and minor mergers alone cannot be the
main mechanism governing the late evolution of elliptical galaxies,
because their simulated merger remnants did not follow the Faber-Jackson
(\citealt{1976ApJ...204..668F}) and Kormendy relations
(\citealt{1977ApJ...218..333K}), although they stayed on the
fundamental plane. Still, dry major and minor mergers can bring
compact early-type galaxies closer to the fundamental plane  but the
size increase estimated from idealized models taking into account the
cosmological context might be too weak
\citep{2009ApJ...706L..86N,2012arXiv1202.0971N,2012arXiv1202.5403C}.   

In this paper we revisit collisionless one- and two-component
equal-mass and minor mergers focusing on the merger dynamics and the
effect of a dark matter halo on the resulting evolution of global
theoretical and observable galaxy properties. In section \ref{NM} we
present the method used to construct equilibrium one- and two-component
systems. The effects of violent relaxation and stripping are discussed
in section \ref{secrelax} followed by a detailed discussion of analytic
estimates for the structural evolution based on the virial theorem in
section \ref{VT}. In section \ref{observ} we present the evolution of
observable galaxy properties. We summarize and discuss the results in
section \ref{summary}.

\section{NUMERICAL METHODS}
\label{NM}

\subsection{Galaxy models}
\label{basics}
For the initial galaxy models we assume spherical 
symmetric, isotropic Hernquist density profiles
(\citealt{1990ApJ...356..359H}) for the luminous as well as the dark matter
 component, 
\begin{equation}
\label{densh}
\rho_{i}(r) = \frac{M_{i}}{2\pi}\frac{a_{i}}{r(r+a_{i})^{3}} ,
\end{equation}
where $\rho_{i},M_{i}$ and $a_{i}$ are the density, the mass and the
scale length of the respective component $i$. The potential is 
\begin{eqnarray}
\label{poth}
\Phi_{i}(r) = -\frac{GM_{i}}{r+a_{i}}, 
\end{eqnarray}
with the gravitational constant $G$. 

On the one hand the projected Hernquist profile is a reasonable
approximation of the $R^{1/4}$ law (\citealt{1948AnAp...11..247D}) 
for the luminous component (its Sersic index however is closer to
$n\sim 2.6$, see \citealt{2006MNRAS.369..625N}). On the other hand it
is a good representation of the \citet{1997ApJ...490..493N} profile
for the dark matter component. Therefore we consider the Hernquist
density distribution a sufficiently realistic description for the
luminous and dark matter distributions of a typical elliptical
galaxy. 

\begin{figure}
  \begin{center}
    \includegraphics[width=8cm,height=12cm]{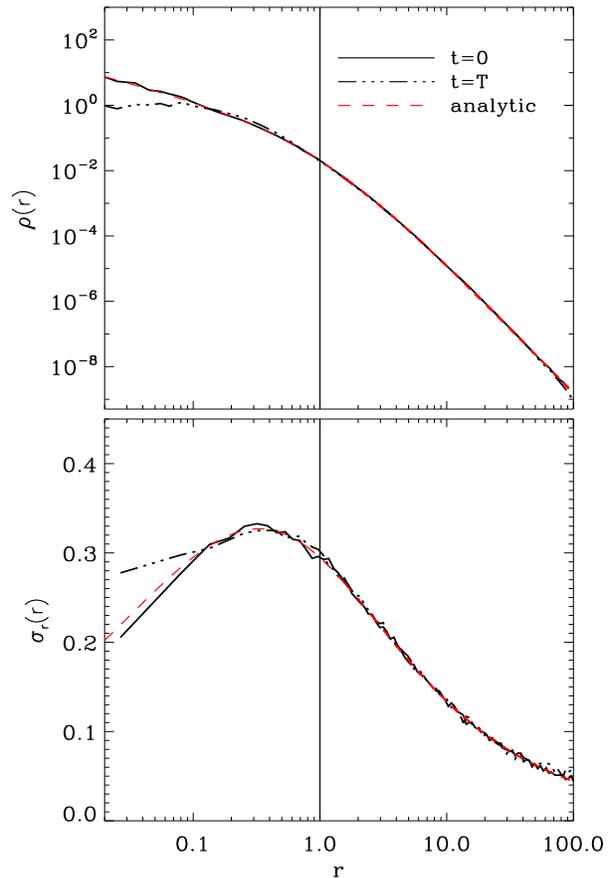}
  \end{center}
  \caption[title]
          {Top panel: Radial density profile for a Hernquist distribution
            (160k particles). The black solid line illustrates the initial 
            profile (t=0) and the dashed-dotted line the final profile 
            ($t=T \sim 100\times t_{dyn}$, where $t_{dyn}$ is 
            the dynamical time at the spherical half-mass radius).The 
            analytical profile is indicated by the red dashed line. 
            Inside 10\% of the scale length (vertical solid line) the
            system is affected (increase in dispersion and decrease in 
            density) by two-body effects. They conduct heat into the 
            initially cooler central regions. However, overall, the system 
            is dynamically stable for at least 100 dynamical times.
            Bottom panel: Initial (solid line) and final (black dashed line) 
            radial velocity dispersion profile.
            
          } 
  \label{fig1}
\end{figure}

For simplicity, we assume isotropy of the velocity distribution to
construct a stable initial configuration, and the bulge and halo
component are in dynamical equilibrium. We compute the distribution
function (DF) $f_{i}$ for each component $i$, using Eddington's
formula \citep{2008gady.book.....B}, 
\begin{equation}
\label{df}
f_{i}(E) = \frac{1}{\sqrt{8}\pi^{2}}\int_{\Phi=0}^{\Phi=E}\frac{d^{2}\rho_{i}}{d\Phi_{T}^{2}} 
\frac{d\Phi_{T}}{\sqrt{E-\Phi_{T}}},
\end{equation}
where $\rho_{i}$ is the density profile of component $i$, $E$ is the
relative (positive) energy and $\Phi_{T}$ is the total gravitational
potential $\Phi_{T}=\Phi_{*}(+\Phi_{dm})$.  Solving distribution
functions, is in general more complicated than using Jeans equations,
but results in more stable initial conditions
\citep{2004ApJ...601...37K}. The distribution function of a two
component Hernquist model can be computed analytically, even in the
anisotropic case \citep{1996ApJ...471...68C}. Our model, however, also
allows for more general density distributions and we calculate $f_{i}$
numerically. As in our approach we have no analytic expression for
$\rho_{i}(\Phi_{T})$ (see Eq. \ref{df}) we have to 
transform the integrand of Eq. \ref{df} to be a function of radius $r$, 
\begin{eqnarray}
\label{trans}
\frac{d^{2}\rho_{i}}{d\Phi_{T}^{2}}d\Phi_{T}& = & \biggl(\frac{d\Phi_{T}}{dr}\biggr)^{-2}
\biggl[\frac{d^{2}\rho_{i}}{dr^{2}}- \nonumber \\
& &-\biggl(\frac{d\Phi_{T}}{dr}\biggr)^{-1}\frac{d^{2}\Phi_{T}}{dr^{2}}\frac{d\rho_{i}}
{dr}\biggr]\frac{d\Phi_{T}}{dr}dr.
\end{eqnarray}
This procedure always results in an analytical expression for the
integrand, even for more general $\gamma$- profiles
\citep{1993MNRAS.265..250D} with different slopes for the density
distribution. As a consequence, the integration limits also have to
change, e.g. $\Phi(r)=0$ gets $r=\infty$ and $\Phi(r)=E$ has to be
solved (numerically) for the radius $r$. 

Once we have computed the DF we can randomly sample particles with
radii smaller than a given cut-off radius and random velocities, which
are smaller than the escape velocity. Then the particle configuration
for the galaxy is established using the Neumann rejection method. 

The one component model is described by two parameters, the scale
length $a_{*}$ and the total mass $M_{*}$. For the two component
models, including dark matter, we additionally introduce the
dimensionless parameters $\mu$ and $\beta$ for the scale length of the
halo $a_{dm}=\beta a_{*}$ and its mass $M_{dm}=\mu M_{*}$.

\subsection{Model Parameters and Merger Orbits}
\label{modpar}

For the total dark matter to stellar mass ratio we assume
$\mu=M_{dm}/M_{*}=10$ and the ratio of the scale radii is $\beta =
a_{dm}/a_{*} = 11$, for all simulations with two-component models. We
perform a set of simulations for two different scenarios; both adopt
$M_{*} = a_{*} = 1.0$. In the major merger scenario we simulate
mergers of initially identical, spherically symmetric one-
or two-component models on zero energy orbits. The encounters have
parabolic orbits with and without angular momentum (head-on). For
higher merger generations we duplicate the merger remnant after
reaching dynamical equilibrium at the center and merge them again on
orbits with the same energy but different infall directions. In total,
we simulate three generations of head-on equal-mass mergers, and two
generations with angular momentum (see also Table \ref{table1}).

\begin{figure*}
  \begin{center}
    \includegraphics[width=15cm]{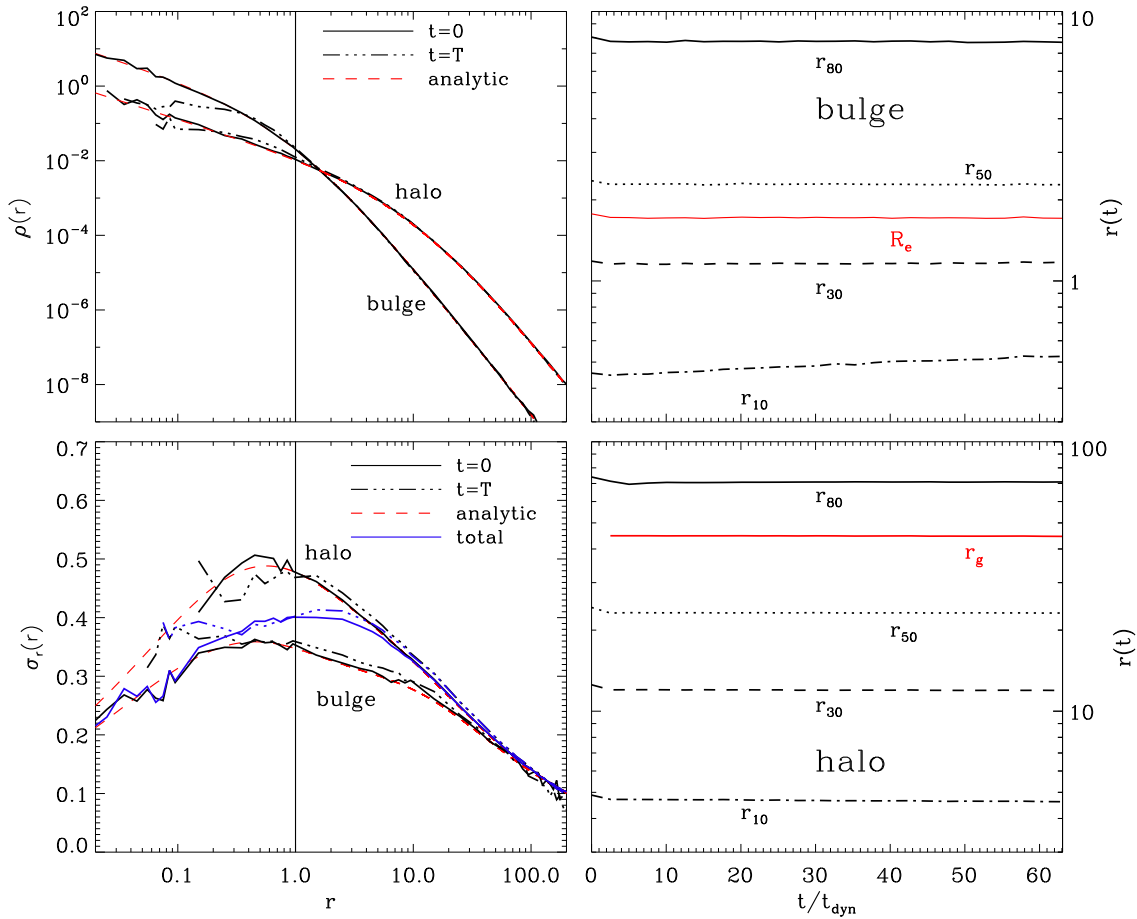}
  \end{center}
  \caption[title]
          {
            Top left panel: Radial density profile for the
            two-component realization with 1100k particles. The bulge
            to halo mass ratio is 1:10. The 
            solid black lines illustrate the initial profiles (t=0) of 
            the bulge and the halo and the dashed-dotted lines their 
            final profiles ($t > 60t_{dyn}$). The analytic Hernquist
            profiles are indicated by the red dashed line. 
            Bottom left panel: Radial velocity dispersion of the total 
            system (blue solid: initially, blue dashed-dotted: final),
            the bulge and  
            the halo separately. Inside $0.3a_{*}$ the model is
            affected by two-body relaxation, but overall it is
            stable. Right panels: Time evolution of the radii
            enclosing 80\%, 50\%, 30\% and 10\%  
            mass (black lines from top to bottom) of the bulge (top
            panel) and halo  
            (bottom panel). The red lines show the effective radius of the 
            bulge (upper panel) and the gravitational radius of the whole system
            (bottom panel). Except for the 10\% radius of the bulge,
            which shows a slight  
            increase, all mass radii stay constant over $> 60$ dynamical times.}
  \label{fig2}
\end{figure*}

In the minor merger scenario our simulation sequences start with an
initial mass ratio of 1:10, i.e. $M_{\mathrm{host}}=10 M_{\mathrm{sat}}$ and a stellar
scale radius of the satellite of $a_{\mathrm{*,sat}}=1.0$. This choice
for the satellite's scale radius seems unrealistically large but there
are no direct observations for low mass galaxies at high redshift and
the sizes of less massive ellipticals at present converge 
towards an effective radius of $r_{e}\sim 1kpc$
\citep{2011MNRAS.414.3699M}. Therefore the satellite galaxies have the
same size as the compact early-type hosts, although they are an order
of magnitude less massive. For comparison, we simulated two sequences (with
one- and two-component models), of head-on minor mergers, where the
satellite's scale radii are half the host's scale radius
$a_{\mathrm{*,sat}}=0.5$, though the satellites lie on an extrapolation of the
observed mass-size relation of \citet{2010ApJ...713..738W} at $z=2$
(see Table \ref{table1}). The host galaxy for the next generation is
the virialized end product of the previous accretion event. This host
is merged with a satellite identical to the first generation. The mass
ratio for this merger is now 1:11. We repeat this procedure until the
host galaxy has doubled its mass, i.e. 10 minor mergers. The final
mass ratio of the merger is 1:19. Again we simulate one- and
two-component mergers with zero (head-on) and non-zero angular
momentum. As the mergers of the bulge+halo model with angular momentum
are computationally expensive we only simulate 6 generations.

For all head-on mergers we separate the centers by a distance $d$ and
assign them a relative velocity $v_{\mathrm{rel}}=2\sqrt{GM_{h}/d}$, where
$M_{h}$ is the total attracting mass of the host galaxy within the
radius d. This velocity corresponds to an orbit with zero energy and
zero angular momentum, i.e. the galaxies will have a zero relative
velocity at infinite distance. The distance $d$ is always large enough
to obtain virialized remnants at the end of each generation. As the
merger remnants after the first generation will not be spherical
anymore, their mutual orientation is randomly assigned at the
beginning of each new merger event. For the mergers with angular
momentum we set the impact parameters to 
half of the spherical half-mass radius of the host's bulge and
separate the galaxies for enough so that the initial overlap is very
small. 

\subsection{Simulations and Stability Tests}
\label{code}

\begin{table}
\begin{tabular}{cccccccccc}\hline\hline
  Run      &  Gen. & $a_{*,sat}$ &  $M_{ub} (\%) $  & $M_{*,ub} (\%)$\\\hline
 B1ho      &  3    &   1.0      &   12.3           &   12.3    \\
 B1am      &  2    &   1.0      &   15.0           &   15.0    \\
 HB1ho     &  3    &   1.0      &   10.1           &    2.5    \\
 HB1am     &  2    &   1.0      &    8.8           &    2.0    \\
 B10hod    &  10   &   1.0      &   21.8           &   21.8    \\
 B10amd    &  10   &   1.0      &   20.9           &   20.9    \\
 B10hoc    &  10   &   0.5      &   21.7           &   21.7    \\
 HB10hod   &  10   &   1.0      &   35.6           &   20.4    \\
 HB10amd   &  6    &   1.0      &   19.5           &    7.9    \\
 HB10hoc   &  10   &   0.5      &   20.9           &    7.2    \\
 \hline
\end{tabular}
\caption
    {This table gives the name of the hierarchy (1st column), the
      number of generations (2nd), the initial scale radius of the
      satellite (3rd), the amount of unbound mass of the total final
      remnant (4th) and the corresponding stellar mass loss (5th). The
      name can be explained as followed; B/HB: bulge or bulge+halo,
      1/10: major/minor merger, am/ho: orbit with/without angular
      momentum. In the case of the minor merger scenarios, c/d
      indicates wether we chose a compact or diffuse satellite. 
    }
\label{table1}
\end{table}

All simulations were performed with VINE \citep{2009ApJS..184..298W,
2009ApJS..184..326N}, an efficient, parallelized tree-code. We use a
spline softening kernel with a softening length $\epsilon=0.02$ for
all runs. In general, the softening length depends on the particle
number
(e.g. \citealt{1996AJ....111.2462M};\citealt{2001MNRAS.324..273D}) and
we found $\epsilon=0.02$ to result in stable models. For the major
merger simulations the host galaxy consists of $N_{*}=1.6\times 10^5$
bulge particles for the one-component (bulge only) model and
$N_{*}=2\times 10^4$ for the two-component model, which has an
additional halo represented by $N_{\mathrm{DM}}=2\times 10^5$ particles of the same
mass. For the accretion scenario, the one- and two-component host
galaxies both have $N_{*}=10^5$ bulge particles and the latter has
$N_{\mathrm{DM}}=10^6$ halo particles. The satellites have ten times fewer
particles for all components.  

In Fig. \ref{fig1} we demonstrate the stability of the bulge only
model with 160k particles by comparing the initial and final (100
dynamical times) density (top panel) and radial velocity dispersion
(bottom panel) as a function of radius to the analytical values. In
general the model is very stable for the full simulation time, except
in the innermost parts, where two-body 
relaxation becomes important. At the highest densities, inside 10$\%$
of the scale radius the relaxation time $t_{\mathrm{relax}}$ of the model is
very small ($t_{\mathrm{relax}} \sim 50 \cdot t_{\mathrm{dyn}}$). Consequently, two-body
encounters change the central particle's energy and deplete the high
density regions. Looking at the initial and final number of particles
within $0.1 a_{*}$, we find that half of the particles escape this
region and go to lower binding energies. However, at larger radii the
models are very stable with a very good agreement to the analytical
solution. The radii enclosing 30, 50 and 80 per cent of the stellar
mass stay perfectly constant. 

The stability of the two-component system is demonstrated in
Fig. \ref{fig2}. Our initial model, constructed of two Hernquist
spheres, is again very stable over a long simulation period of more
than 60 dynamical times (Figure 2). The density and the velocity
dispersion do not change significantly. Again the innermost regions
are affected by two-body relaxation. However, looking at the mass radii
of the bulge and the halo we observe no big changes. The apparent
contraction of the mass radii enclosing 80$\%$ or 50$\%$ is less than
5$\%$ for the halo and less than 2$\%$ for the bulge. The
gravitational radius and the effective radius which we use in this
paper stay constant. Therefore we conclude that the regions we are
interested in are not affected by  
two-body relaxation and other numerical effects of the initial
conditions, and the results of our merger simulations should be robust
for our choice of force resolution and particle numbers.

\section{Relaxation \& Stripping}
\label{secrelax}

Our set of simulations indicate that dissipationless mergers of spheroids are 
strongly affected by two dynamical processes, {\it violent relaxation}
which dominates the evolution of major mergers and {\it tidal
  stripping} which is important for minor mergers. In the following we
will discuss both processes in more detail. 

\begin{figure*}
  \begin{center}
    \includegraphics[width=16cm]{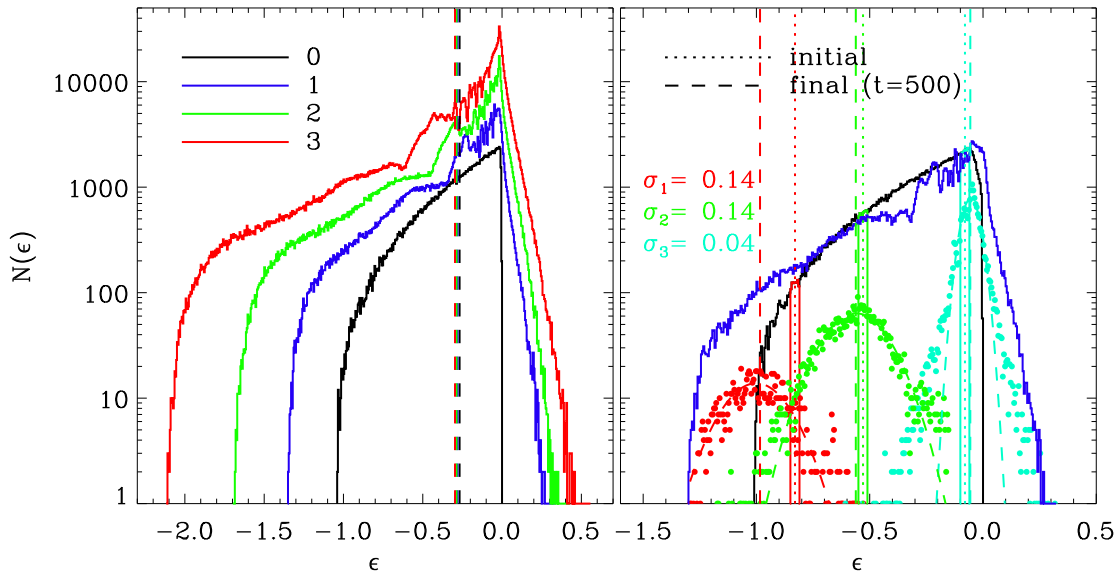}
  \end{center}
  \caption[title]
          {
            Left panel: Energy distribution for the initial
            one-component host galaxy and three generations of head-on
            major mergers (B1ho). The initial distribution (solid
            black line) is broadened during each merger towards 
            lower and higher (escapers) energies. Finally the most
            bound particles have about two times their initial  
            binding energy. The mean energy of the total system stays
            constant (vertical dashed lines). Right panel: Energy
            distribution for all particles of the progenitor galaxy
            (black solid line, same as in the left panel) and the same
            particles after one equal-mass merger (blue solid
            line). Particles at high (red solid,
            $-0.85<\epsilon<-0.81$), intermediate (green solid,
            $-0.55<\epsilon<-0.51$) and low initial binding energies 
            (light blue solid, $-0.1<\epsilon<-0.06$) are highlighted and
            their energy distribution after the mergers are shown by
            the dotted distributions. By violent relaxation the two
            innermost bins are broadened by violent relaxation to
            Gaussian distributions, 
            with a width of $\sigma=0.14$. Their  
            mean energies are shifted to lower values (short- and
            dashed vertical lines). The most weakly bound particles 
            are shifted to
            higher energies resulting in a significant fraction of
            escapers.
          }
          \label{fig3}
\end{figure*}

\begin{figure*}
  \begin{center}
    \includegraphics[width=16cm]{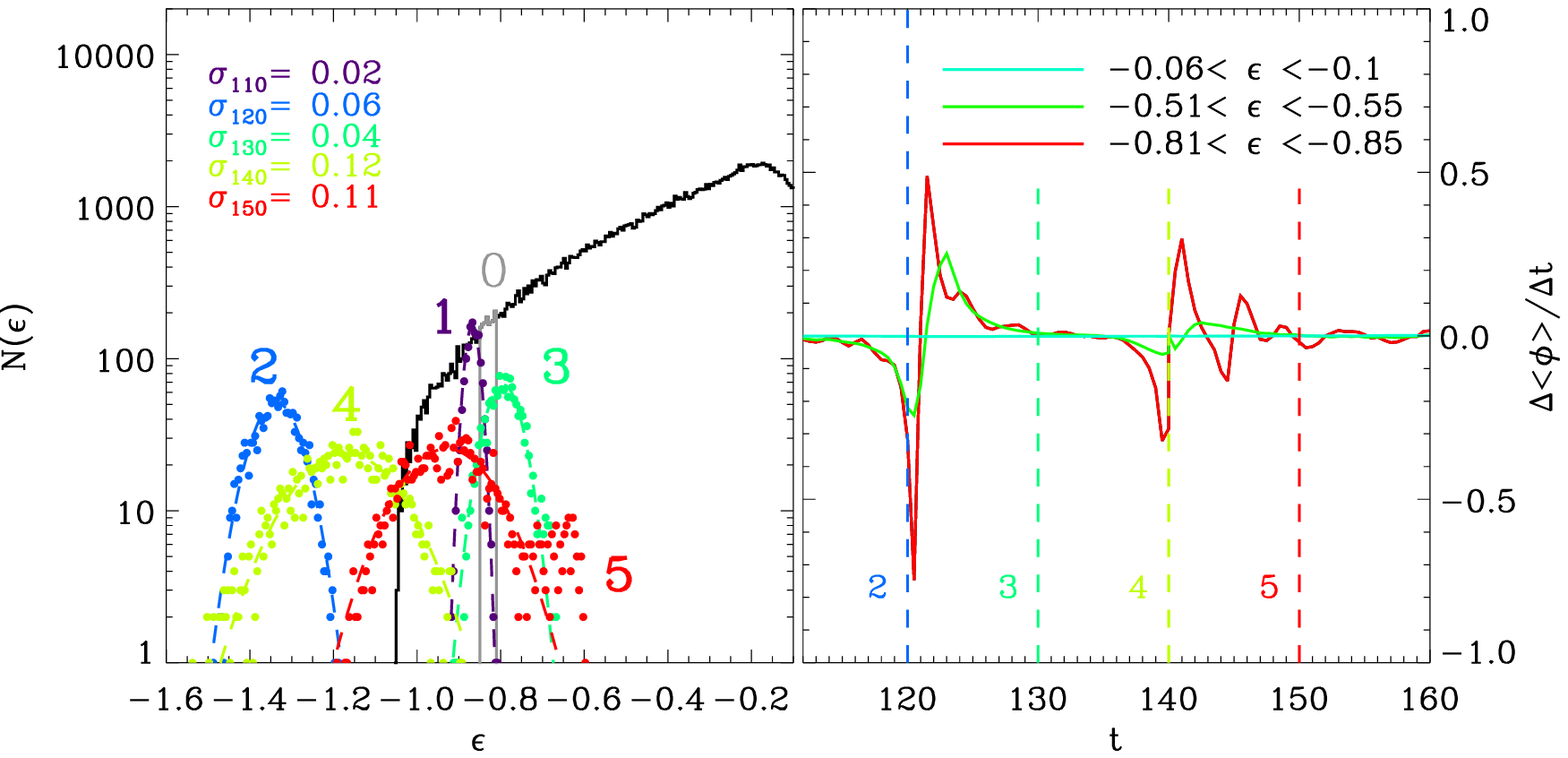}
  \end{center}
  \caption[title]
          {
            Left panel: Temporal evolution of the energy distribution of
            highly bound particles (red bin in Fig. \ref{fig3}). The
            corresponding time 
            evolution during the merger of the average potential of
            this energy bin (red line) and the two others presented in
            Fig. \ref{fig3}. The most bound bin shows the strongest
            fluctuation during the first (2, t = 120) and second (4, t
            = 140) encounter. The initial distribution
            at $t=100$ (0) is slightly broadened to $\sigma_{110}=0.02$ (1) by two-body 
            relaxation. At the first close encounter $t\approx 120$
            (2) the potential rapidly gets deeper (see also right
            panel), the particles are shifted to higher  
            binding energies and the energy distribution widens to $\sigma_{120}=0.06$.
            As the two galaxies fly away from each other, the potential increases 
            and the particles go back to lower binding energies (3) without further
            broadening of the distribution. During the second close
            encounter, the process is repeated, i.e. the distribution
            broadens when the potential gets deeper (4) and moves back
            to higher energies. After $t=150$ (5) the central regions show only negligible
            potential fluctuations and the particle distribution is slowly
            affected by two-body relaxation. During the merger, the innermost energy bin 
            is broadened from $\sigma = 0.02$ to $\sigma = 0.1$ by
            violent relaxation. In isolation, this energy bin is only
            moderately affected by two-body relaxation over the same
            time interval ($\sigma=0.02\rightarrow 0.03$).
          }
          \label{fig4}
\end{figure*}

\subsection{Major Mergers}
\label{vrelax}

For equal-mass mergers, the effect of violent relaxation is very strong
and has a significant effect on the differential energy distribution of the remnants.
The left panel of Fig. \ref{fig3} indicates, that the initial narrow distribution 
(black line) becomes broader with each generation. Tightly bound particles
become even more bound and some weakly bound particles gain enough
energy to escape the galactic potential. The theoretical framework of
violent relaxation is very complex and, since the pioneering work of
\citet{1967MNRAS.136..101L}, there have been many approaches to
obtain a viable theory, which can describe the final equilibrium configuration 
of a violently relaxing system (e.g. \citealt{1978ApJ...225...83S,2000ApJ...531..739N,2005MNRAS.362..252A}).
In the following we first give a short introduction to the original approach of 
\citet{1967MNRAS.136..101L}, before we discuss our results with respect to another
slightly different approach of \citet{1992ApJ...397L..75S}.

During the approach and interaction of two collisionless systems the total 
gravitational potential $\Phi$ varies with time, which leads to a 
non-conservation of energy of single particles 
\citep{1967MNRAS.136..101L,1992ApJ...397L..75S},
\begin{eqnarray}
\frac{d\epsilon}{dt}=-\frac{\partial \Phi}{\partial t},
\end{eqnarray}
where $\epsilon$ is the energy per unit mass. In accordance with the time dependent 
virial theorem,
\begin{eqnarray}
\frac{1}{2}\frac{d^2I}{dt^2}=2T+V,
\end{eqnarray}
where $I$ is the moment of inertia tensor, the galaxy will convert its
total potential energy $V$ into kinetic energy $T$ and back. In
equilibrium, $\ddot{I}=0$ so $T=-E, V=2E$, with $E=T+V$ being the
total energy. Away from equilibrium the total energy $E$ is constant,
but $T$ and $V$ will vibrate about these values, which will widen the
differential energy distribution $N(E)$, where $N(E)$ gives the number
$N$ of stars within an energy interval $E+dE$. 

This evolution is illustrated in the left panel of Fig. \ref{fig3},
where the energy distribution $N(\epsilon)$ broadens with each merger
generation very similar to the results reported in
\citet{1992ApJ...397L..75S}. Using the Ansatz, that violent relaxation
can be approximated by scattering effects of single particles, they
found an analytic prediction for the final equilibrium configuration
for the energy distribution of an equal-mass mergers where the
scattering probability functions become Gaussian. In the right panel
of Fig. \ref{fig3} we select particles in three different bins
with low ($-0.06<\epsilon <-0.1$), intermediate $(-0.55<\epsilon <-0.51)$ and high
$(-0.85<\epsilon <-0.81)$ initial binding energies. After the final merger,
the initially narrow low energy bins are significantly broader and the
two most bound bins can be well fitted with a Gaussian of width
$\sigma=0.14$. Bins at lower energies does not become a Gaussian,
eventually due to escaping particles. The mean energies of tightly bound
particles (red vertical lines) are shifted to even higher binding
energies. The temporal evolution of the latter energy bin is detailed
in the left panel of  Fig. \ref{fig4}. Until $t = 100$ there is little
evolution since the galaxy centers are not yet interacting. Then, at
the first encounter, the mean binding energy increases strongly and
the distribution broadens by a factor of 3 ($\sigma= 0.02\rightarrow
0.06$). When two galaxies fly apart the mean energy nearly reaches its
original value without further broadening ($t=130$). During the second
close encounter, this scenario repeats, i.e. the energy distribution
is shifted to higher binding energies accompanied by a strong broadening
($t=130\rightarrow 140$), before it oscillates back into a less bound
state ($t=140\rightarrow 150$). Now the particles reside at slightly 
higher than initial binding energies. In the right panel of
Fig. \ref{fig4} we depict the evolution of the mean potentials of the
three energy bins in Fig. \ref{fig3},  which oscillate strongly for
the tightly bound particles (red, green lines) and the energy shifts
and broadenings are obviously correlated to the potential
fluctuations. On the other hand these fluctuations vanish rapidly, due
to phase mixing resulting in incomplete relaxation
\citep{1967MNRAS.136..101L,1978ApJ...225...83S}. Violent relaxation
offers new energy states by deepening the total gravitational
potential, into which the most bound particles are scattered making
them even more bound. At the same time a fraction of the weakly bound
particles reach positive energies and escape the potential of the remnant.  
Two-body relaxation effects for the initial system, evolved in
isolation, are much weaker (e.g. 
$\sigma=0.02\rightarrow 0.03$ between $t=110\rightarrow 160$).

\begin{figure}
  \begin{center}
    \includegraphics[width=8.5cm]{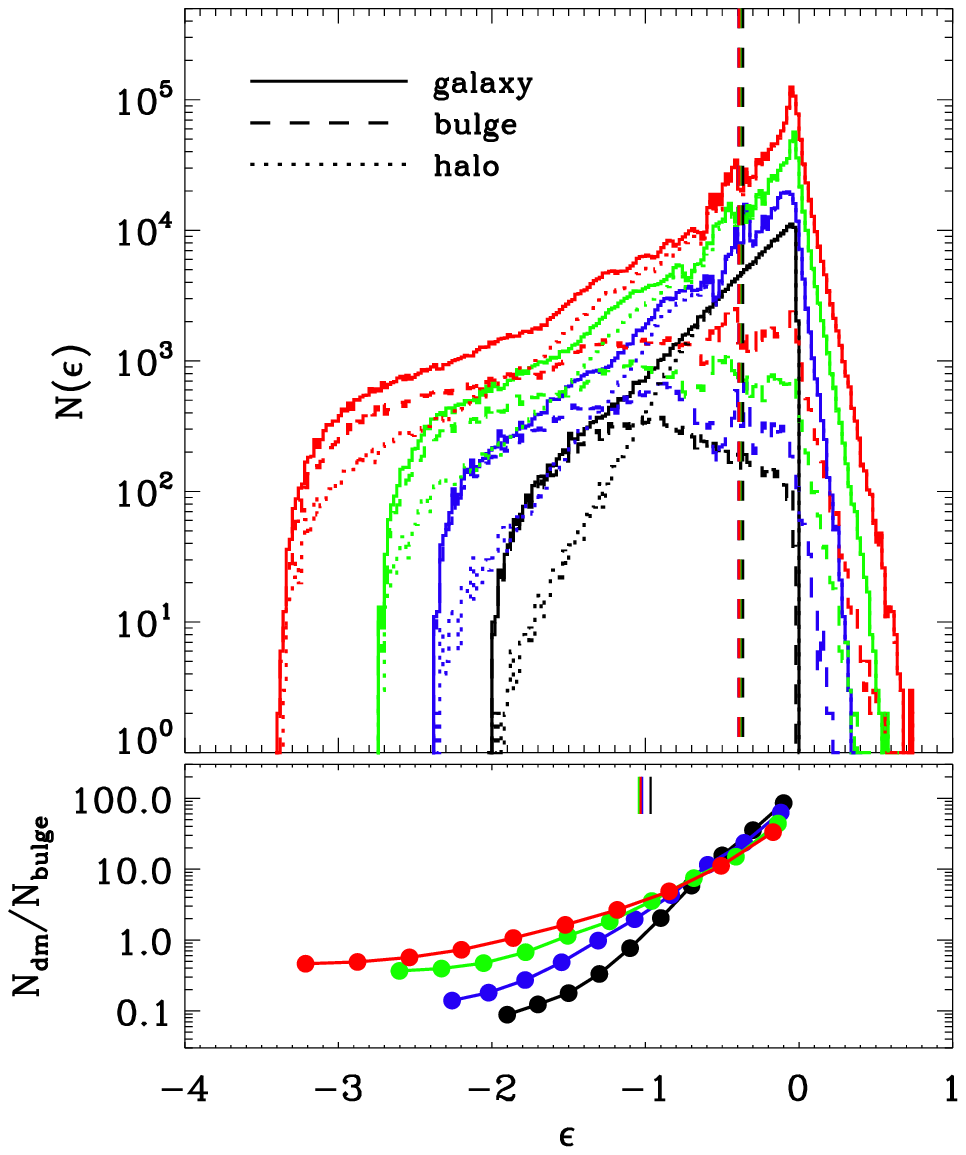}
  \end{center}
  \caption[title]
          {
            Top panel:
            Energy distribution for the two-component initial
            condition (black) and after the first (blue), second
            (green), and third (red) generation of equal-mass mergers. The solid lines
            depict the distribution for all particles, the dashed lines
            for the bulge particles and the dotted lines of the halo
            particles. The general evolution of the total system is similar to the
            one-component system (left panel, Fig. \ref{fig3}). The mean binding
            energy of the systems (vertical dashed lines) stays 
            constant. However, relatively more dark matter particles
            than bulge particles are scattered to low energies (or
            higher binding energies), thereby changing the shape of
            the dark matter energy distribution and increasing the
            central dark matter fraction (see also Fig. \ref{dmfrac}).
            Bottom panel: The ratio of dark matter to bulge particles
            (or mass as we have equal-mass particles) increases with
            each merger generation for energies lower than $\approx
            -0.8$ and decreases at higher energies (only bound
            particles are shown). Finally, there are almost as many
            dark matter than bulge particles in the innermost
            bins. The small vertical lines at the top of the panel  
            indicate the energy including 50\% of the most bound bulge
            particles, which stays constant after the first
            merger.
          }
          \label{fig5}
\end{figure}

\begin{figure}
  \begin{center}
    \includegraphics[width=8.5cm]{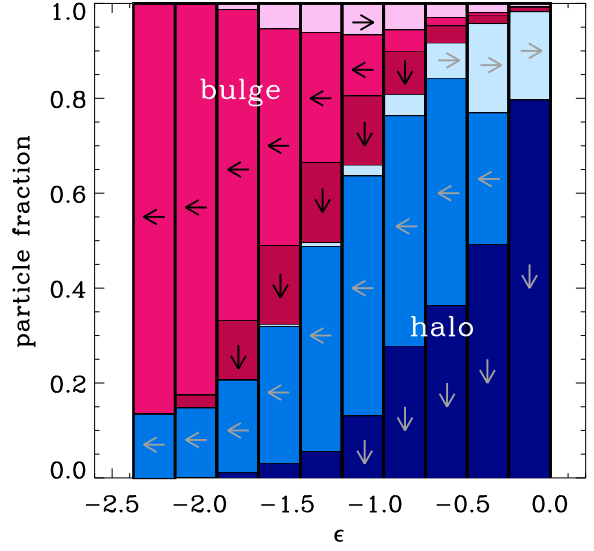}
  \end{center}
  \caption[title]
          {
           This panel indicates the energetic origin of bulge and halo particles
           after the first merger (black to blue line in Fig. 
           \ref{fig5}) in terms of particle fractions per energy
           bin. The lowest energies $\epsilon<-2.0$ were not  
           occupied in the initial system but all particles are
           scattered down from higher energies during the merger
           (indicated by $\leftarrow$ and intermediate 
           blue for halo and intermediate red for bulge particles). Energies $\epsilon\ge -2.0$
           were already initially populated with particles and some of
           them stay in their initial energy bin (indicated by $\downarrow$ and dark
           colors). The light blue/red regions
           show particles scattered up from lower to higher energies 
           (indicated by $\rightarrow$). We
           predominantly observe downflow in regions where the energy
           distribution has a positive gradients, i.e. mostly at high
           binding energies (Fig. \ref{fig5}). This process is
           driven by violent relaxation and scatters more dark matter
           than bulge particles into the center increases (bottom panel of 
           Fig. \ref{fig5}).  Upflow is seen for 
           regions with negative gradients in the energy distribution,
          most prominently at the lowest energy bins leading to
           escapers.  
          }
          \label{fig6}
\end{figure}
In the top panel of Fig. \ref{fig5} we demonstrate, that the overall
energy distribution for major mergers of two-component models
(solid lines) evolves similar to one-component models, i.e. the
tightly bound particles go to states with even higher binding energies
and some weakly bound particles become unbound (positive energies). However,
with each generation, the number of dark matter particles in the
highly bound regions increases more than the number of bulge
particles. This behavior can also be seen in the bottom panel of
Fig. \ref{fig5}. The fraction of dark matter to bulge particles
converges to unity in the central regions. As the energy of the 50\%
most bound bulge particles,  indicated by the small vertical lines at
the top of this panel, hardly evolves the intrinsic structure of the
system has to change. For example there are only very few dark matter 
particles in the most bound regions initially, but a non-negligible amount
occupies the most bound state of the remnant. This implies that the
system undergoes a 'real' change in terms of its dark matter
fraction. The fact, that more dark matter than bulge particles wander
to higher binding energies is illustrated in Fig. \ref{fig6} for the
first merger of HB1ho. The amount of dark matter scattered to more bound states
is larger than the luminous matter for energies $\epsilon \ge -1.1$,
which is the region where the halo starts to dominate the mass (see
top panel Fig. \ref{fig5}). 

This demonstrates that violent relaxation rearranges the distributions
of dark and luminous matter in energy space, which yields a higher
dark matter fraction at the center of the final system. The observable
consequences will be discussed in more detail in section \ref{secdm}. 
 
\subsection{Minor Mergers}

\begin{figure}
  \begin{center}
    \includegraphics[width=8cm]{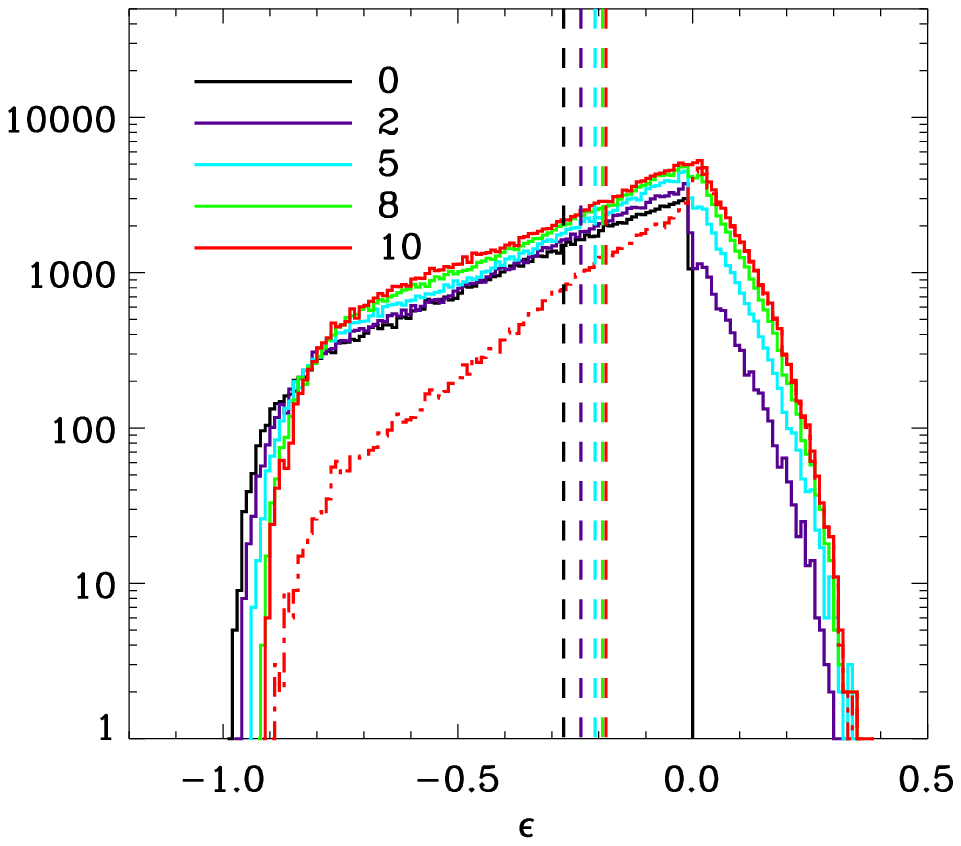}
  \end{center}
  \caption[title]
          {
            Differential energy distribution for the initial
            one-component system (black) and two (purple), five
            (blue), eight (green), and 10 generations of 1:10 head-on 
            mergers (B10hoc). The
            red dashed-dotted line indicates the energy distribution of all
            accreted material, which shows that nearly all escapers come
            from the satellites and nearly no particles assemble at the
            center. In contrast to
            equal-mass mergers (left panel, Fig. \ref{fig4}) all bound particles
            become less bound as the mean binding energy
            of the systems (vertical dashed lines) decreases with each
            merger generation.
          }
          \label{hist_ac}
\end{figure}

\begin{figure}
  \begin{center}
    \includegraphics[width=8cm]{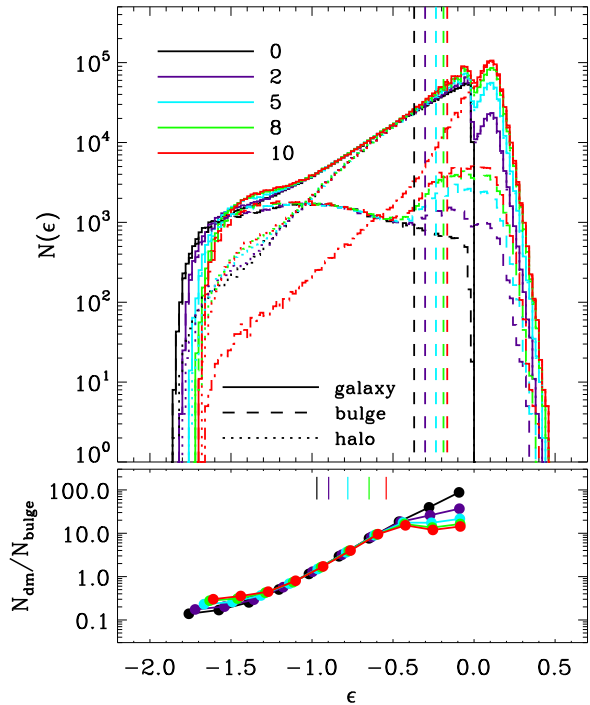}
  \end{center}
  \caption[title]
          {
            Top panel:
            Energy distribution for the initial
            two-component system HB10hod (black) and two (purple), five
            (blue), eight (green), and 10 (red) generations of 1:10
            mergers. The solid lines
            depict the distribution for all particles, the dashed lines
            for the bulge particles and the dotted lines of the halo
            particles. The red dashed-dotted line indicates the energy
            distribution of all satellite particles. For minor mergers
            violent relaxation only unbinds satellite particles and the
            energy distribution of the host is almost unaffected. In
            contrast to 
            equal-mass mergers (Fig. \ref{fig5}) bound particles
            become less bound, due to two-body effects.
            Similar to one-component minor mergers
            (Fig. \ref{hist_ac}), the mean binding energy of the
            systems (vertical dashed lines) 
            decreases with each merger generation.
            Bottom panel:
            In stark contrast to equal-mass mergers (Fig. \ref{fig5}) the
            ratio of dark matter to bulge  
            particles at energies $\epsilon < -0.4$ is unaffected. At higher
            energies the fraction of (stripped) bulge particles
            increases. The short vertical lines  indicate the energy including 50\%
            of the most bound bulge particles. In contrast to equal
            mass mergers this energy constantly
            increases, i.e. the bulge grows into the halo such that
            the enclosed dark mass increases. 
          }
          \label{hist_ac2c}
\end{figure}

In Fig. \ref{hist_ac} and the top panel of Fig. \ref{hist_ac2c} we
show the total energy distributions (solid lines) for a sequence of
head-on, one- and two-component minor 
mergers, respectively. In both cases nearly all escaping particles originate
from the satellites (red dashed-dotted line in both figures), which
indicates that violent relaxation only affects the in-falling 
material and has negligible effects on the distribution of the host
galaxy particles. As the satellites are less bound than the host galaxy we find a very high 
fraction of unbound mass for the final remnants. Furthermore, we can see 
that almost no accreted particles assemble in the central regions. The
loss of binding energy of the most bound particles in both scenarios 
is caused by two-body effects, which conduct heat into the central high 
density regions (see also Fig. \ref{fig1}). Combining this shift with the 
effect, that most satellite 
particles assemble at low binding energies, results in an increase of the
mean binding energies. For the bulge only mergers, the decrease of the 
mean energy can also be predicted analytically.

The potential energy for a Hernquist sphere is
 \citep{1990ApJ...356..359H},
\begin{eqnarray}
W=-\frac{GM^2}{6a},
\end{eqnarray}
and according to the virial theorem the total energy of a system in equilibrium is
\begin{eqnarray}
E = \frac{1}{2}W=-\frac{GM^2}{12a}.
\end{eqnarray}
Additionally we define the mass ratio of accreted to initial host mass
$\eta\equiv\frac{M_a}{M_i}$  and the ratio of the according 
scale radii as $\zeta\equiv\frac{a_a}{a_i}$. 
Assuming energy conservation for a zero energy orbit, the system's final energy is:
\begin{eqnarray}
E_f &=& E_i+E_a=-\frac{M_i^2}{12a_i}-\frac{M_a^2}{12a_a}  \\
    &=& -\frac{M_i^2}{12a_i}-\frac{(\eta M_i)^2}{12\zeta a_i} = -\frac{M_i^2}{12a_i}(1+\frac{\eta^2}{\zeta})  \\
    &=& E_i(1+\frac{\eta^2}{\zeta})
\end{eqnarray}
Furthermore we calculate the mean final energy $\varepsilon$,
\begin{eqnarray}
\varepsilon_{f} &=& \frac{E_{f}}{M}=\frac{E_i(1+\frac{\eta^2}{\zeta})}{M_{h}+M_{s}} = \frac{-\frac{M_i^2}{12a_i}(1+\frac{\eta^2}{\zeta})}{M_{h}(1+\eta)} \\
                &=& -\frac{M_i}{12a_i}\biggl(\frac{1+\frac{\eta^2}{\zeta}}{1+\eta}\biggr)=-\varepsilon_i\biggl(\frac{1+\frac{\eta^2}{\zeta}}{1+\eta}\biggr),
\end{eqnarray}
where we used the fact that for equal mass particles the total number of particles
is equivalent to the total mass $M$.
In the case of an equal mass merger of two identical systems $\eta=\zeta=1$ 
and $\varepsilon_f=\varepsilon_i$, i.e. the mean energy of the system stays 
constant (see also Figs. \ref{fig3}, \ref{fig5}). But for the numerical 
setup of the
first minor merger (B10hoc), where $\eta=0.1$ and  $\zeta=1$, the final mean
energy is  $\varepsilon_f=0.927\cdot\varepsilon_i$, in agreement with
the simulations (Fig.\ref{hist_ac}). 

Taking a closer look on the energy distribution of the bulge of the scenario HB10hod
(dashed lines, top panel Fig. \ref{hist_ac2c}) we can directly see, that most stellar 
particles accrete at energies $\epsilon >-0.4$, creating an overdensity of bulge 
particles. Consequently the ratio of dark matter particles to bulge particles 'decreases'
for $\epsilon>-0.4$ (bottom panel of Fig. \ref{hist_ac2c}). While this
ratio stays constant for all particles with $\epsilon<-0.4$ the
lowest binding energy of the 50\% most bound particles moves to higher
energies and the dark  matter fraction increases (see also Fig.\ref{dmfrac}). Altogether we can see, that
the dark matter halos of the satellites are dissolved very rapidly in the deep potential
well of the host galaxy and efficient tidal stripping leads to a build up of an stellar 
overdensity at low binding energies.

Comparing the results of minor mergers with either compact or diffuse
satellites, the overall evolution stays the same for one- and
two-component models. But, as the compact satellites have a higher
binding energy, tidal stripping is less efficient and the particles
assemble closer to the center.  However, the very innermost regions are
still hardly affected. 

\section{Virial Theory}
\label{VT}
Going back to virial theory, \citet{2009ApJ...699L.178N} presented a simple 
prediction of how stellar systems evolve during a merger event, which
will be extended later on and therefore is repeated briefly.  
Using the virial theorem and assuming energy conservation we can 
approximate the ratios of the initial to the final mean square speed 
$\langle v_{\mathrm{i/f}}^2 \rangle$, 
gravitational radius $r_{\mathrm{g,i/f}}$ and density $\rho_{\mathrm{i/f}}$ of a merging system.
According to \citet{2008gady.book.....B} the total energy of a system is

\begin{figure*}
  \begin{center}
    \includegraphics[width=18cm]{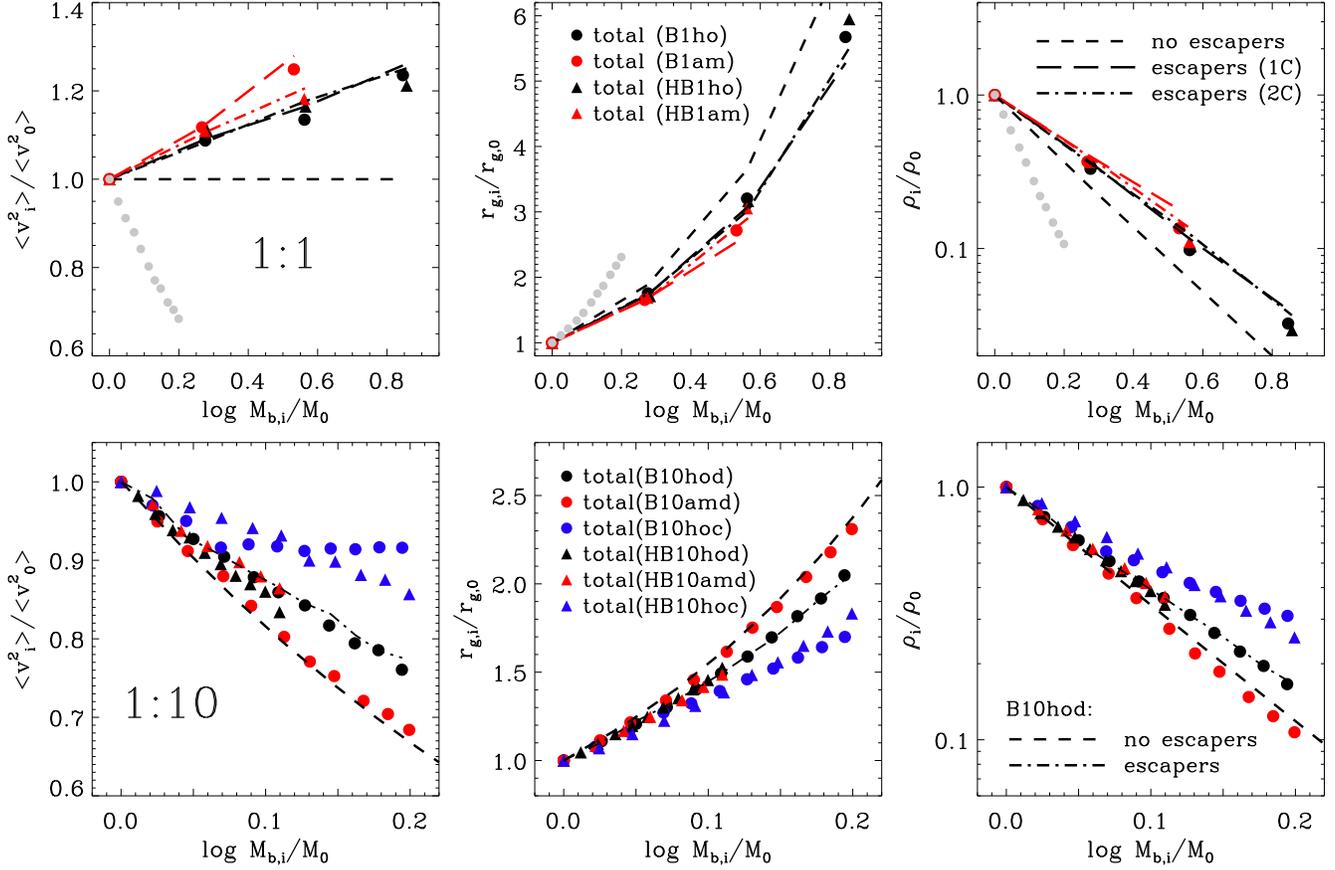}
  \end{center}
  \caption[title]
          {
            Top panels: Evolution of the mean square speeds (left), the 
            gravitational radii $r_{g}$ (middle) and the spherical density within
            $r_{g}$ (right) for the equal mass merger generations. The circles 
            (triangles) represent the evolution of the
            one- (two-) component models with (red) and without (black) angular momentum.
            The black short dashed lines indicate the prediction of Eqs. 
            \ref{thorsi1}-\ref{thorsi3}. Considering the escaping particles Eqs.
            \ref{new1}-\ref{new3} yield the long dashed lines and the dashed-dotted lines
            for the one- and two- component scenarios, respectively. 
            For better comparison, the light grey circles highlight 
            the evolution of the minor merger scenario B10amd.
            Bottom panels: The symbols show the evolution of the minor merger scenarios
            (see Table \ref{table1}) for the one- (circles) and two- (triangles) component
            models with (red) and without (black) angular momentum. The blue circles
            depict the results, using compact satellite galaxies. The 
            dashed lines are the idealized expectations of \citet{2009ApJ...699L.178N}
            and the dashed dotted lines are accounting for the escapers in the 
            scenario B10hod.
          }
  \label{rg_new}
\end{figure*}

\begin{eqnarray}
  \label{btenergy}
    E_{i} &=& K_{i}+W_{i} = -K_{i} = \frac{1}{2}W_{i}\nonumber  \\
    &=&-\frac{1}{2}M_{i}\langle v^{2}_{i}\rangle = -\frac{1}{2}\frac{GM^{2}_{i}}{r_{g,i}},
\end{eqnarray}
where $E_i$ and $M_i$ are the system's initial total energy and mass. The gravitational
radius is defined as
\begin{eqnarray}
\label{rg}
r_{g,i} \equiv \frac{GM_i^2}{W_i},
\end{eqnarray}
with the total potential energy $W_i$. Now we define $E_{a}$, $M_{a}$, $r_{\mathrm{g,a}}$ and 
$\langle v^2_{a}\rangle$ as the energy, mass, gravitational radius and mean square speed
of the accreted system. Additionally to the former defined mass ratio $\eta=M_a/M_i$,  
$\epsilon=\langle v^{2}_{a}\rangle/ \langle v^{2}_{i}\rangle$ is the dimensionless velocity 
fraction. By combining these assumptions with equation \ref{btenergy} we obtain
\begin{eqnarray}
\label{thorsi1}
\frac{\langle v^{2}_{f}\rangle}{\langle v^{2}_{i}\rangle}=\frac{(1+\eta\epsilon)}{(1+\eta)},\\
\label{thorsi2}
\frac{r_{g,f}}{r_{g,i}}=\frac{(1+\eta)^2}{(1+\eta\epsilon)},\\
\label{thorsi3}
\frac{\rho_{f}}{\rho_{i}}=\frac{(1+\eta\epsilon)^3}{(1+\eta)^5},
\end{eqnarray}
for the ratios of the final to initial mean square speed, gravitational radius and density.
In the very simple case of an equal mass merger of two identical systems $\eta=\epsilon=1$, 
$r_{g}$ gets doubled (Eq. \ref{thorsi2}), $\langle v^2 \rangle$ stays constant (Eq. 
\ref{thorsi1}) and $\rho$ decreases by a factor of 4 (Eq. \ref{thorsi3}). 
If we use equations 7-9 for a minor merger scenario, where $\langle v^2_a \rangle 
<< \langle v^2_i \rangle$ and $\epsilon << 1$, the size of the final system can increase by 
a factor of $\sim$ 4, as $r \propto M^{2}$. Additionally, the final velocity dispersion and 
density are reduced by a factor of 2 and 32, respectively.
These changes are quantitatively in good agreement with observations. 

However, this simple analytic model suffers from a number of limitations, apart from the
restrictions to parabolic orbits and collisionless systems. The effect of violent
relaxation \citep{1967MNRAS.136..101L} in the rapidly changing potential during the merger
will scatter particles in energy space, making some more bound and unbinding others 
from the system, thus energy is not perfectly conserved (see Section \ref{secrelax}). 
Additionally, realistic spheroidal galaxies are composed of two collisionless
components, dark and luminous matter with different spatial distributions, which are 
expected to react differently to a merger event \citep{2006MNRAS.369..625N,
2009ApJ...691.1424H}. 

\subsection{Major Mergers}
\label{secrgmaj}

Starting with the major mergers, the top left panel of Fig. \ref{rg_new}
shows the evolution of the mean square speeds of the
one- and two-component equal-mass mergers with the total bound mass of
the system. According to Eq. \ref{thorsi1} the mean square speed
should remain unchanged (dashed line), but obviously it increases with
each generation. As a consequence, the growth of the total
gravitational radius (top middle panel of Fig. \ref{rg_new}) and the
density decrease (top right panel of Fig. \ref{rg_new}) are weaker than
expected. The same trend was reported by
\citet{2003MNRAS.342..501N,2009ApJ...706L..86N} who correctly argued that the 
simple analytical prediction is only valid for an idealized case
without escaping particles. In the 4th column of Table \ref{table1} we
show, that the amount of unbound mass after the merger is not  
negligible and adds up to about 12, 15, 10 and 9\% per cent of the 
total mass for the scenarios B1ho, B1am, HB1ho and HB1am, respectively.
The same effect can be seen for the two-component case, where two
merger generations with angular momentum 
have nearly as much mass loss as three generations of the head-on counterparts
($\approx 9\%$ compared to $\approx 10\%$). Looking at the last column of
Table \ref{table1} we can see that nearly all escaping particles for
the two-component models are from the halo, as nearly no stellar mass
gets lost ($M_{\mathrm{*,ub}}<3\%$). 

We can now revisit the analytic predictions and
take the effect of escapers into account (see also
\citealp{2003MNRAS.342..501N}). The energy 
equation using the energy of the bound final system $E_{f}$ and the
energy of the escaping particles $E_{\mathrm{esc}}$ is  
\begin{eqnarray}
  E_{f}+E_{esc} = E_{i} + E_{a}.
\end{eqnarray}
We assume that the escaping particles have essentially zero potential energy, so that
\begin{eqnarray}
  E_{esc}=+\frac{1}{2}M_{esc}\langle v^2_{esc}\rangle.
\end{eqnarray}
With $\alpha \equiv M_{\mathrm{esc}}/M_{i}$ as the ratio of mass, lost in
escapers to initial mass and $\beta \equiv \langle v^{2}_{\mathrm{esc}}\rangle/
\langle v^{2}_{i}\rangle$  as the ratio of the mean square speed  
of the escapers and the initial system, we can now re-write equations
\ref{thorsi1} to \ref{thorsi3} as  
\begin{eqnarray}
\label{new1}
\frac{\langle v^{2}_{f}\rangle}{\langle v^{2}_{i}\rangle}=\frac{(1+\eta\epsilon+\alpha\beta)}{(1+\eta-\alpha)}, \\
\label{new2}
\frac{r_{g,f}}{r_{g,i}}=\frac{(1+\eta-\alpha)^2}{(1+\eta\epsilon+\alpha\beta)} 
\end{eqnarray}
and
\begin{eqnarray}
\label{new3}
\frac{\rho_{f}}{\rho_{i}}=\frac{(1+\eta\epsilon+\alpha\beta)^3}{(1+\eta-\alpha)^5}.
\end{eqnarray}

\begin{figure}
  \begin{center}
    \includegraphics[width=8cm]{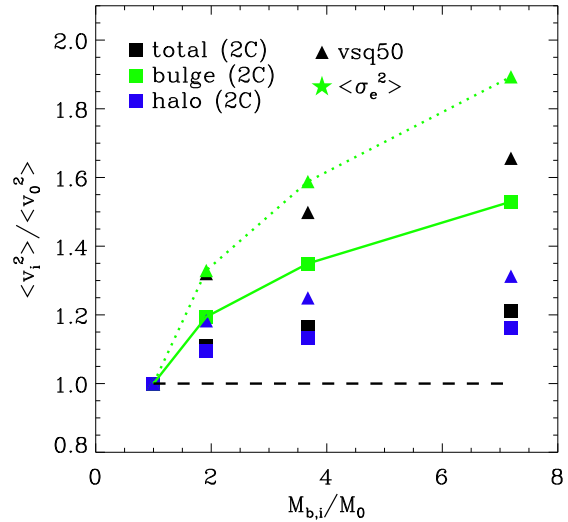}
  \end{center}
  \caption[title]
          {
            The evolution of the mean square speeds of the total system 
            (black squares), the halo (blue squares) and the bulge
            (green squares) with bound bulge mass for 
            two-component models on head-on orbits (HB1ho). 
            The more rapid increase of the bulge velocities can be explained by 
            an energy transfer from the halo to the bulge (green solid line, 
            see Eq. \ref{deltav}). At the center this effect is 
            even more efficient, as the mean square speeds within the spherical 
            half-mass radius (corresponding triangles) of the bulge 
            increase more. The dotted green line indicates the expectation of 
            Eq. \ref{deltav} for the central region.
          }
          \label{mass_mm}
\end{figure}

The long dashed (one-component models) and dashed-dotted (two-component models)
lines in the top panels of Fig. \ref{rg_new} indicate that the updated analytic 
predictions are in good agreement with our simulation
results. The deviations are less than a few per cent for both models.

The situation becomes more complicated if we separate the velocities of
the bulge and the halo component (Fig. \ref{mass_mm}). The mean square 
speed of the bulge (green
squares) increases more (finally $> 50\%$) with respect to the total
system (black squares), whereas the halo (blue squares) speed stays
below the total. Here violent relaxation and dynamical friction lead
to an energy transfer from the bulge to the halo (see section
\ref{secrelax}), i.e. the final bulge is more tightly bound than the
initial one (see also \citealt{2005MNRAS.362..184B} for a discussion
of the effect of different orbits).  

This effect can be estimated based on the ratio of dark and stellar
matter. The total kinetic energy of the system is
\begin{eqnarray}
  m_{*}\langle v^{2}_{*}\rangle+m_{dm}\langle v^{2}_{dm}\rangle=m_{tot}\langle v^2_{tot} \rangle.
\end{eqnarray}
With $m_{\mathrm{tot}} \equiv m_{*}+m_{\mathrm{dm}}$ and introducing $\Delta \langle v^2_{\mathrm{*/dm}} \rangle =
\langle v^2_{\mathrm{*/dm}} \rangle - \langle v^2_{\mathrm{tot}} \rangle$ we obtain, 
\begin{eqnarray}
   m_{*}\Delta\langle v^{2}_{*}\rangle+m_{dm}\Delta\langle v^{2}_{dm}\rangle=0
\end{eqnarray}
and the additional growth of the stellar mean square speeds is
\begin{eqnarray}
  \Delta \langle v^{2}_{*}\rangle =-\frac{m_{dm}}{m_{*}}\Delta\langle v^{2}_{dm}\rangle.
  \label{deltav}
\end{eqnarray}
If we now add $\Delta \langle v^2_{*} \rangle$ to the mean square speed of the galaxy 
$\langle v^2_{\mathrm{tot}} \rangle$ we can consistently predict the bulge
dispersion (green solid line in Fig. \ref{mass_mm}).

\begin{figure}
  \begin{center}
    \includegraphics[width=8cm]{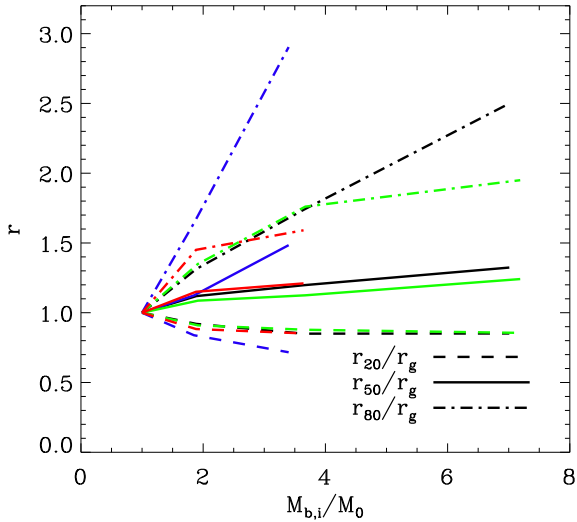}
  \end{center}
  \caption[title]
          {
            Evolution of the radii enclosing the 20\% (dashed), 50\%
            (solid) and 80\% (dashed-dotted) most bound particles normalized to
            the evolution of the gravitational radius , $r_{g}$ (black: B1ho; blue: B1am;
            green: HB1ho; blue: HB1am). The inner regions expand less and the 
            outer regions more than the gravitational radius. All
            ratios are normalized to the initial values.
          }
  \label{contr}
\end{figure}

As violent relaxation scatters some particles into states with higher
binding energy, the central region contracts slightly
relative to the total system growth (depicted by the gravitational radius).
In Fig. \ref{contr} we show the radii enclosing the
20, 50 and 80 \% most bound particles normalized to the evolution of
the gravitational radius. The inner regions expand less, the outer
regions more than the gravitational radius. This effect is consistent
with the analysis in section \ref{secrelax} and was already
described in \citet{1978MNRAS.184..185W,1979MNRAS.189..831W}. It leads
to a significant non-homology of the system (see also
\citealt{2005MNRAS.362..184B}), as it evolves through successive
mergers.    

On the other hand, this high central density  leads to higher central velocities 
(triangles, Fig. \ref{mass_mm}). If we now add $\Delta \langle v^2_{*} 
\rangle$ (Eq. \ref{deltav}) to $\langle v^2_{\mathrm{tot}} \rangle$ for the central regions 
we again get a very good prediction of the central bulge velocity (green dotted 
line Fig. \ref{mass_mm}).

\subsection{Minor Mergers}
\label{secrgmin}

The bottom left panel of Fig. \ref{rg_new} indicates that the mean square speeds 
of all minor merger hierarchies decrease with increasing mass. In all 
scenarios with diffuse satellites (black and red symbols), the evolution is very 
close to the virial expectations of 
Eqs. \ref{thorsi1}-\ref{thorsi3} (dashed line), although the mass loss is significant,
especially for the two-component models (red and black triangles). In Table \ref{table1} we 
can see that the fraction of escaping particles is up to $35\%$ for HB10hod and around
$20\%$ for all other minor merger scenarios. Furthermore, regarding the 2C models, 
most of the escape
fraction is due to the dark matter particles. Going back to the evolution of 
$\langle v^{2}\rangle$,
we can see that the corrected prediction of Eq. \ref{new1} (dashed-dotted line), 
which includes the effect of mass loss, perfectly fits the results (e.g. scenario B10hod). 
Using more compact satellites the final decrease of velocities (blue symbols)
is much weaker, because they are more tightly bound. As they have half the scale radius of
the diffuse satellites, their binding energies and velocities are two times higher
which then doubles the velocity fraction $\epsilon=\langle v^{2}_{a}\rangle/ \langle v^{2}_{i}\rangle$ of Eqs. \ref{thorsi1}-\ref{thorsi3} and yields a smaller decrease. In combination
with the occurring mass loss, this explains the different evolution of the mean square speeds.
Nevertheless, in all scenarios the final mean square speeds of the total systems are 
$10-30\%$ lower compared to their initial host galaxies.

The evolution of the gravitational radii (bottom middle panel, Fig. \ref{rg_new}) 
of the six hierarchies evolve according to the mean square speeds, which
is not surprising as $r_{g}\propto 1/\langle v^{2}\rangle$ (see Eq. \ref{btenergy}). 
In detail, this means, that the hierarchies with a diffuse satellite show a size increase,
which is consistent with the analytic predictions of Eq. \ref{thorsi2} (dashed line). 
As the compact satellites are not able to efficiently decrease the mean square speeds, 
the gravitational radii grow only marginally.
However, for all minor mergers the maximum size growth is around a
factor $\sim 2.4$. For completeness, the bottom right panel
illustrates, that the mean density within the gravitational radius
evolves according to the gravitational radius ($\rho\propto
r_{g}^{-3}$), i.e. $\rho$ decreases at maximum by 90\% (model B10amd,
red circles). 

\section{Observables}
\label{observ}

\subsection{Size \& Dispersion Evolution}
\label{obsre}

\begin{figure*}
  \begin{center}
    \includegraphics[width=18cm]{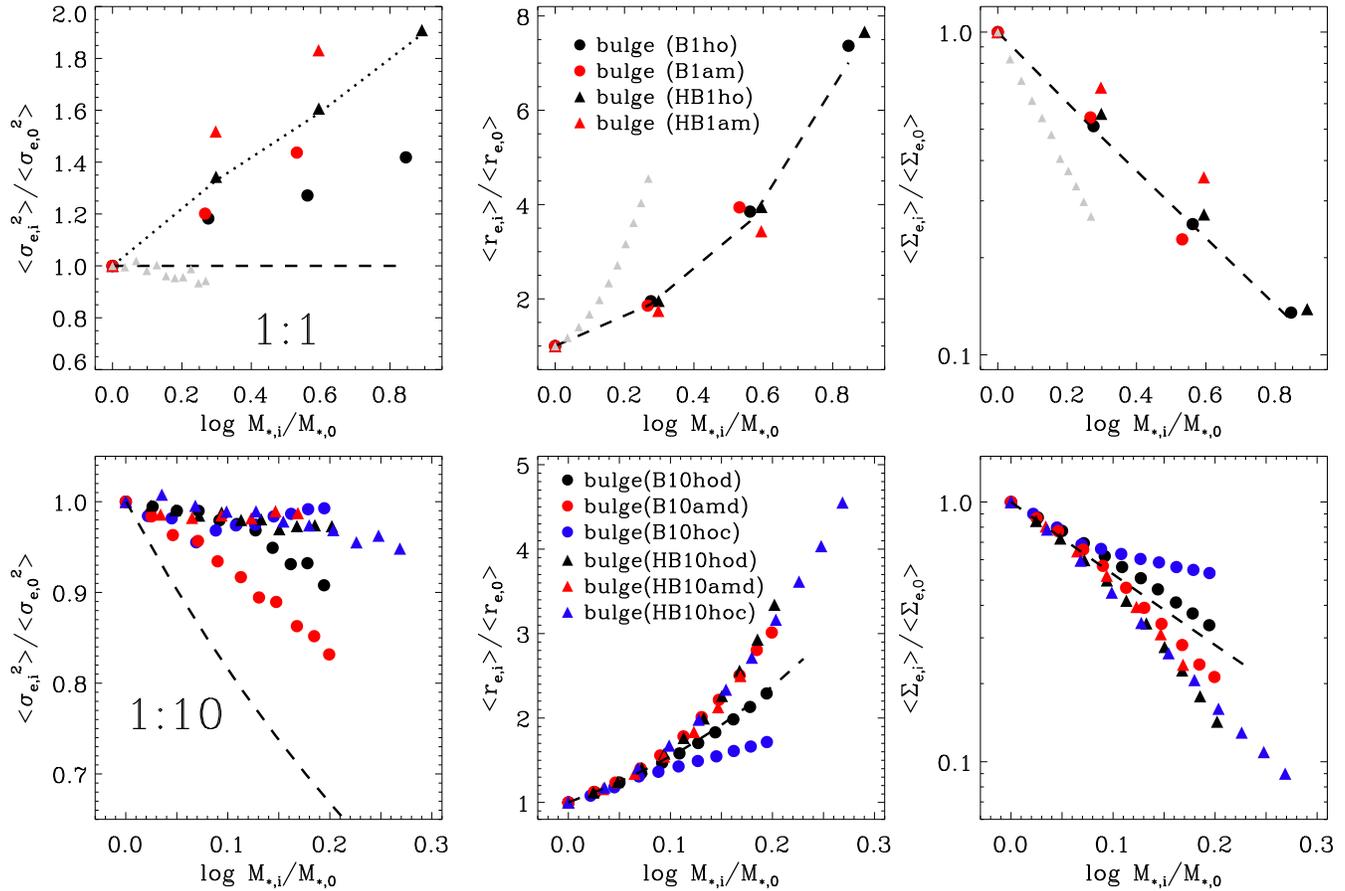}
  \end{center}
  \caption[title]
          {
           Same as Fig. \ref{rg_new} but for 'observable' bulge properties like
           the effective line-of-sight velocity dispersions (left panels), 
           the projected half-mass (effective) radii (middle panels) and
           the effective surface densities (right panels) 
           versus the bound stellar mass normalized to the initial
           stellar mass. The top panels show the evolution of the
           major merger and the bottom panels of the 
           minor merger hierarchies, respectively. The light grey 
           triangles in the top panels are the results of the minor merger
           scenario HB10hoc.
           The dashed lines indicate the simple analytic
           predictions (Eqs. \ref{thorsi1}, \ref{thorsi2}, and
           \ref{thorsi3}) neglecting escapers and energy transfer.
           The dotted line in the top left panel accounts for both effects
           fits perfectly $\sigma_{e}^2$ for scenario HB1ho (see also
           Fig. \ref{mass_mm}).  
          }
  \label{reff_new}
\end{figure*}

Now we switch focus from theoretical galaxy properties, like
the gravitational radii, to directly observable galaxy properties. We
compute the effective radius, $r_{e}$, as the mean radius including
half of the projected bound stellar mass along the three principal
axes, the mean projected stellar velocity dispersion, $\sigma_{e}$,
within this radius and $\Sigma_{e}$ as the mean projected mass surface
density within $r_{e}$. 

\subsubsection{Major Mergers}

The black triangles in
the top left panel of Fig. \ref{reff_new} indicate, that the effective
velocity dispersion $\sigma_{e}^2$ for the two-component mergers with
head-on orbits (HB1ho) evolves in a similar fashion to the central
mean square speeds of the bulge (green triangles Fig. \ref{mass_mm}),
with a final value a factor $\sim 1.9$ higher than initially. In the
case of one-component models, the central 
region of the remnants just suffer from the contraction effect
(Fig. \ref{contr}) and $\sigma_{e}^2$ is only slightly higher than the
mean square speed of the total system (top left panel
Fig. \ref{rg_new}). This effect is stronger for the scenario with
angular momentum orbits. Here, $\sigma_{e}^2$ increases more 
compared to the head-on case (top left panel Fig. \ref{reff_new}) and
the escapers change the dispersions of both systems with an additional
transfer of kinetic energy from the halo to the bulge particles.
(see section \ref{VT}).

\begin{figure}
  \begin{center}
    \includegraphics[width=8cm]{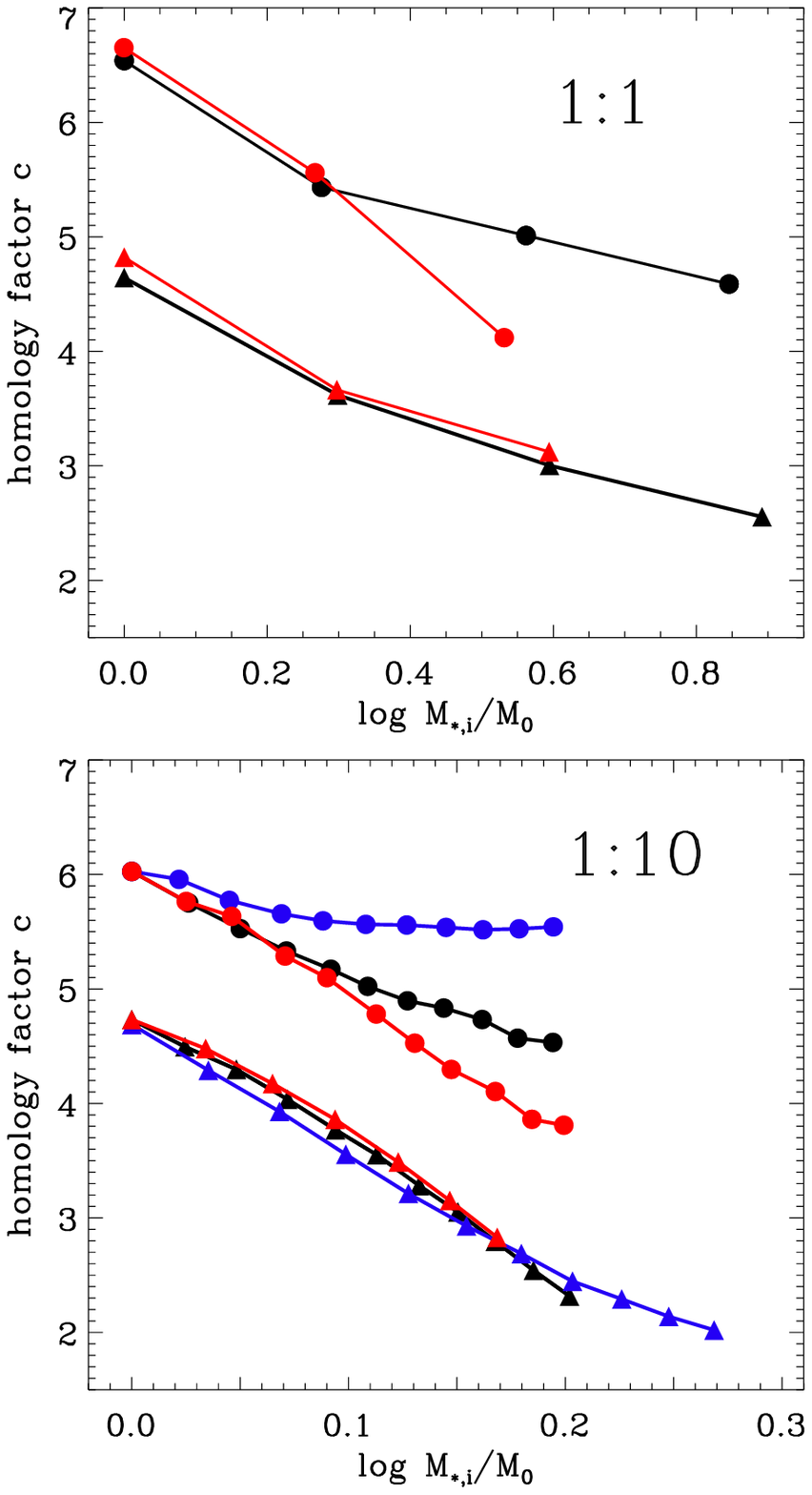}
  \end{center}
  \caption[title]
          {
            Evolution of the structure parameter $c$ relating the
            stellar mass with the effective radius and velocity
            dispersion $ c = (r_{\mathrm{e}} \sigma^2)/(M_* G)$ with bound
            stellar mass for 1:1 mergers (top panel) and 1:10 mergers
            (bottom panel). For all cases $c$ decreases and homology is
            not preserved. In particular the structure of the
            two-component minor mergers (triangles) changes 
            rapidly with added stellar mass. The colors and symbols in
            both panels are the same as in Fig. \ref{rg_new}.
          }
  \label{hom_acc}
\end{figure}

In contrast to the gravitational radii, the observable effective radii
of one- and two-component major mergers (top middle panel Fig. \ref{reff_new})
follow the simple analytic predictions of Eqs. \ref{thorsi1}-\ref{thorsi3}, i.e. the
effective radius grows proportional to the mass, $r_{e} \propto M^{\sim 1.0}$.
At first this seems surprising as from the virial theorem
$r_{e} \propto M/\sigma_{e}^2$ and we would expect smaller radii as
$\sigma_{e}^2$ increases. Projection effects can be ruled out, as the
spherical half-mass radii of the 
bound remnants evolve similar to $r_{e}$ and it cannot be an effect of
dark matter alone. 

As discussed in \citet{2003MNRAS.342..501N} and
\citet{2005MNRAS.362..184B} we also find, that the systems change
their internal structure with each merger generation. This
non-homology effect can be quantified by the structure parameter $c$
which connects the stellar mass of the systems to its observed size
and velocity dispersion
\begin{eqnarray}
M_{*}=c\cdot \frac{r_{e} \sigma_{e}^2}{G}.
\label{hom}
\end{eqnarray}
We use $M_{\mathrm{*}}$ as the bound bulge mass of the remnants to get
a comparable value for $c$ of both merger hierarchies (see also
\citealt{1997A&A...321..111P, 2009ApJ...703.1531N}). A change in $c$
indicates that the merger 
remnants do not have a self-similar structure. For the major mergers 
we find a continuous decrease of $c$ with each merger generation  
(top panel Fig. \ref{hom_acc}). The decrease is stronger (a factor of 1.8) 
for the models including dark matter (triangles) than for the bulge only models
(factor of 1.5; circles), which can be explained by an
increasing central dark matter fraction. In section \ref{vrelax} we have 
already shown, that this is not just an effect of increasing radii (as
argued by \citealt{2009ApJ...703.1531N}), but due to violent
relaxation. Our results just confirm the findings for major mergers
presented in \citet{2005MNRAS.362..184B}. The top right panel of Fig. \ref{reff_new}
depicts the stellar surface densities, which evolve according to
$r_{e}$ and decrease by almost an order of magnitude for a stellar mass increase of a
factor $\sim 8$. For the size increase with mass 
 \begin{eqnarray}
 r_{e} \propto M_{*}^{\alpha},
 \end{eqnarray}
we find $\alpha = 0.8-1.0$ for the first equal-mass merger generation, similar to  
\citet{2005MNRAS.362..184B} for comparable merger orbits.  The
dispersion evolves as 
 \begin{eqnarray}
 M_{*} \propto \sigma_{e}^{\beta},
 \end{eqnarray}
 with $\beta = 3.3-5.1$. 

\subsubsection{Minor Mergers}

In the bottom panels of Fig. \ref{reff_new} we show the evolution of
$\sigma_{e}$ (left), $r_{e}$ (middle) and $\Sigma_{e}$ (right) for all
minor mergers. The stellar dispersion $\sigma_{e}^2$ evolves only
weakly and does not show the theoretically expected decrease, except
for the two bulge-only scenarios with diffuse satellites 
(B10amd, B10hod). 

\begin{figure}
  \begin{center}
    \includegraphics[width=8cm]{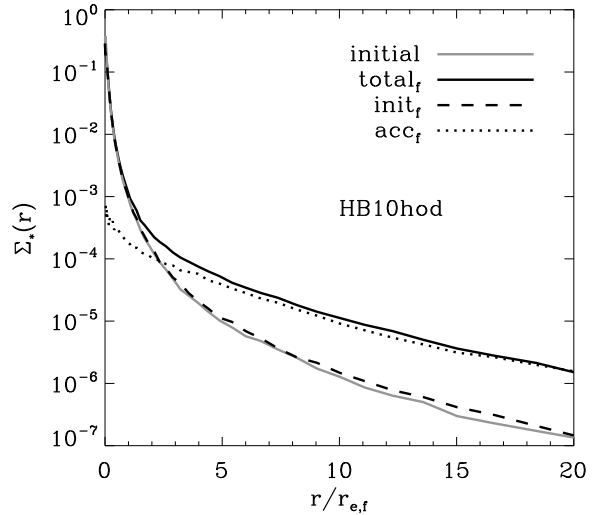}
  \end{center}
  \caption[title]
          {
            Surface mass densities of bulge particles along the major
            axis, normalized to the final effective radius for head-on minor mergers (model HB10hod) of
            two-component models. The grey solid line indicates the  
            initial host stellar surface density, which almost remains
            unchanged (black dashed line). Most of the satellite stars (dotted line)
            assemble with a shallower surface density profile and
            dominate the outer parts of the total surface density of
            the remnant at $ r > r_{\mathrm{e}}$ (black solid
            line). The accreted stars are responsible for the rapid
            size growth. 
          }
  \label{surf_acc2c}
\end{figure}

On the other hand the effective radii $r_{e}$ (middle panel) grow 
significantly and in particular for the two-component models the final
size can be a factor of $4.5$ larger for a mass increase of a about a
factor of two. This is more than the maximum value from simple virial
expectation (Eq. \ref{thorsi2}). For all minor merger scenarios of 
two-component models we get a mass evolution of $r_{e} \propto M^{\sim
  2.4}$, similar to \citet{2012arXiv1202.0971N} for comparable
orbits. This rapid increase is caused by stripped satellite particles
which assemble at large radii (see section \ref{secrelax}). As an
example we show the final stellar surface density distribution of
model HB10nod in Fig. \ref{surf_acc2c}. The host particles are
concentrated at the center similar to their initial distribution. The
accreted stellar particles have a much shallower surface density distribution
and build up an envelope which dominates at radii $r \geq
r_{\mathrm{e}}$. The structure parameter evolves strongly and very similar for all
two-component systems. Its evolution mainly reflects the rapid size
growth (black, red and blue triangles in the bottom panel of
Fig. \ref{hom_acc}). This strong evolution is also seen for the
effective surface densities (bottom right panel of
Fig. \ref{surf_acc2c}), which can decrease at maximum by an order of
magnitude for a size increase of a factor of two. 

For bulge-only minor mergers scenarios the
structural changes depend on the initial conditions. Minor mergers
with compact satellites (B10hoc) evolve nearly self-similar with only
mild changes in the structural parameter (bottom panel of
Fig. \ref{hom_acc}) and a weak evolution in $\sigma_{e}$, $r_{e}$ and
$\Sigma_{e}$ (bottom panels of Fig. \ref{reff_new}).  The observable
parameters evolve very similar to the theoretical ones (bottom panels
of Fig. \ref{rg_new}). The other two bulge only minor merger models
have weakly bound satellites, which already loose most of their material 
in the outer regions of the host galaxy and also build up an extended 
envelope with a corresponding change in structural properties  (see
also black and red circles in Fig. \ref{hom_acc}). 
On the other 
hand, the development of an extended envelope boosts the size growth of a system. As the
sequence B10amd with an angular momentum orbit needs more time until 
the final coalescence, it suffers more from tidal stripping and builds up the most extended 
envelope of all bulge only models, which then results in the highest size growth 
(red circles, Fig. \ref{reff_new}). This 
implies, that the calculation of $\sigma_{e}$ also includes particles outside the innermost
regions, where the velocities are lower and the velocity dispersion within the effective 
radius decreases.

\subsection{Dark Matter Fractions}
\label{secdm}

 \begin{figure}
  \begin{center}
    \includegraphics[width=8cm]{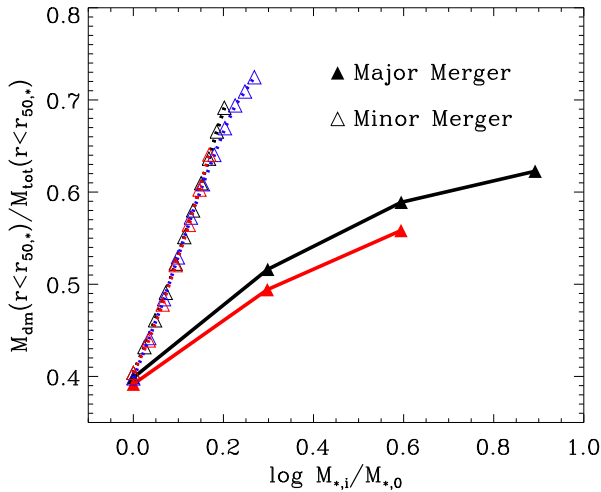}
  \end{center}
  \caption[title]
          {
            The dark matter fraction within the spherical half-mass
            radius of the bulge, $f_{\mathrm{dm}} =
            M_{\mathrm{dm}}/(M_{\mathrm{dm}}+M_{*})$,  
            versus the bound stellar mass of the two-component major (filled triangles) 
            and minor (open triangles) merger remnants. The colors are the same as in 
            Fig. \ref{rg_new}. For major mergers, the dark matter
            fraction increases by $50\%$ for a mass increase of a
            factor $\sim 8$ ($25\%$ for a mass increase of a factor
            $\sim 2$). Due to the rapid increase of the effective  
            radii, the dark matter fraction of the minor mergers
            increases by almost $80\%$ for a mass increase of less
            than a factor $2$.  
          }
  \label{dmfrac}
\end{figure}

As already discussed in section \ref{secrelax} mergers of bulge-halo
systems in general increase the central dark matter fractions. For
major mergers this process is driven by violent relaxation changing
the composition of the remnants  whereas in minor mergers structure is 
almost unchanged but the 'ruler' which is the size of the stellar
system covers larger radii with initially higher dark matter
contributions. The 'observable' quantity here 
is the dark matter fraction within the stellar
half-mass radius which is presented in Fig. \ref{dmfrac} (see also \citealt{2011MNRAS.415.2215B} ). For major
mergers the dark matter fraction increases with every merger
generation, probably saturating for later generations. A stellar mass increase
of a factor two results in an increase of the dark matter fractions by
$\sim 25\%$ for the first generation and $\sim 18\%$  and $\sim 5\%$
for the second and third generation, respectively. For minor mergers
the increase in dark matter fraction is almost linear to the mass
increase and significantly more dramatic than for major mergers. A
mass increase of a factor two results in an increase of the dark
matter fraction by $\sim 80\%$, which is three times more than for
major mergers, without any signature for saturation. This implies that
merging, minor mergers in particular, is a significant driver for the
increase of dark matter in galaxies which had to be more baryon
dominated in the past if their assembly history was dominated by
stellar merging. We note that for minor mergers the increase in dark
matter fraction seems to be roughly proportional to the mass growth.

\subsection{Velocity Dispersion Profiles}

\begin{figure}
  \begin{center}
    \includegraphics[width=8cm]{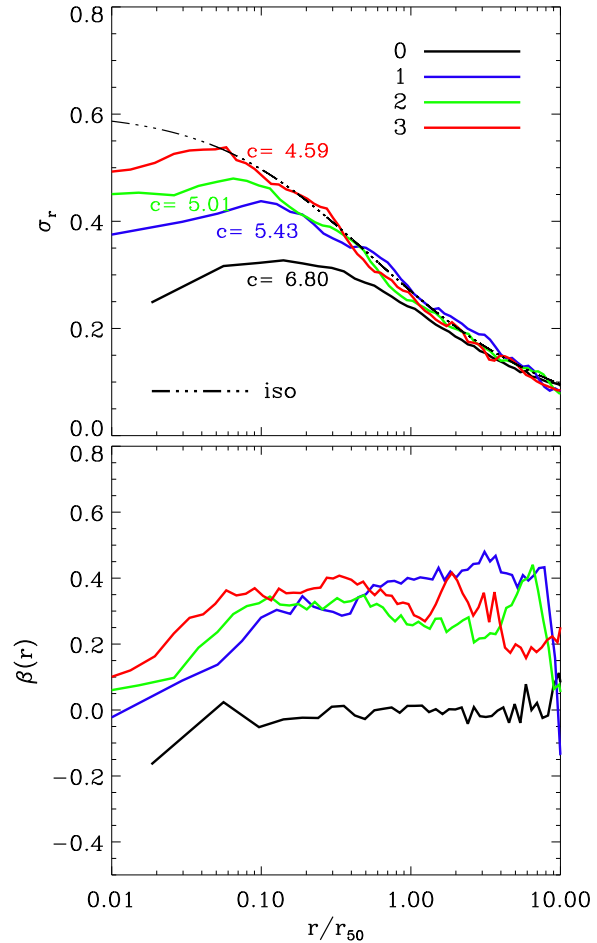}
  \end{center}
  \caption[title]
          {
            Top: Final radial velocity dispersions of the one-component merger 
            generations (B1ho). The colors depict the different generations. The 
            dashed-dotted 
            line indicates the velocity dispersion of a Jaffe profile, which has the 
            same scale length and mass as the last remnant. The latter
            profile resembles the inner parts of the singular isothermal sphere which 
            is a very good fit to our last merger remnant. As the structure parameter
            $c$ decreases with each generation homology is not preserved.
            Bottom: Anisotropy parameter $\beta$ (eq. \ref{aniso}) against radius of three 
            generations of one-component equal-mass mergers. As $\beta>0$ 
            for higher generations the remnants become radially anisotropic over the
            whole radial range. The half-mass radius $r_{50}$ is the radius
            of the sphere, which includes half of the bound system mass.
          }
          \label{sigr1c}
\end{figure}

\begin{figure}
  \begin{center}
    \includegraphics[width=8cm]{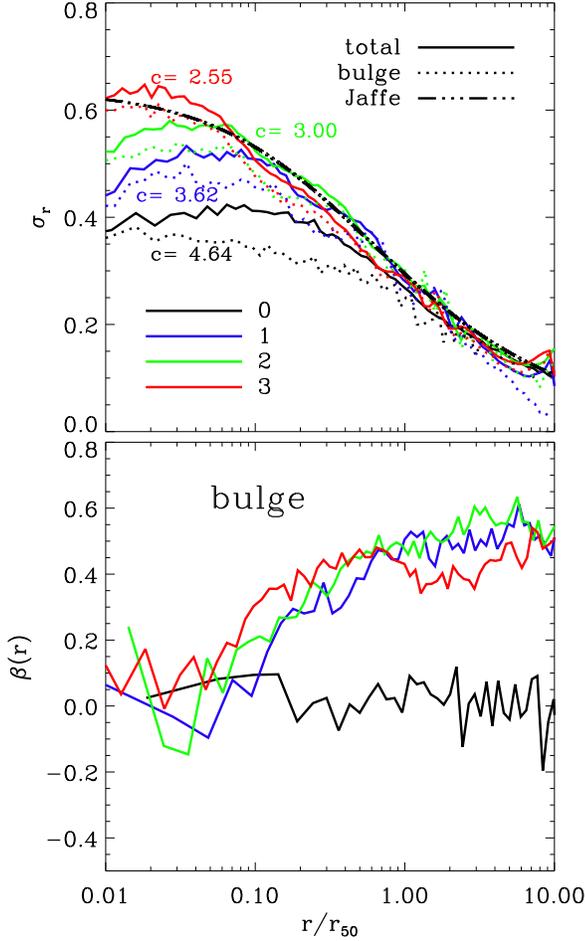}
  \end{center}
  \caption[title]
          {
            The top panel shows the same as in Fig. \ref{sigr1c} for the total 
            remnants (solid lines) and the bound bulges (dotted lines) of 
            the head-on two-component mergers (HB1ho). The bottom panel depicts
            the anisotropy parameter for the bulge. Here, $r_{50}$ is the
            spherical half mass radii of the total (top) and stellar (bottom)
            bound remnants.
            If the bulge is embedded in a dark matter halo, $\beta$ is slightly
            higher compared to the one-component case.
          }
          \label{sigr1d}
\end{figure}

\begin{figure}
  \begin{center}
    \includegraphics[width=8cm]{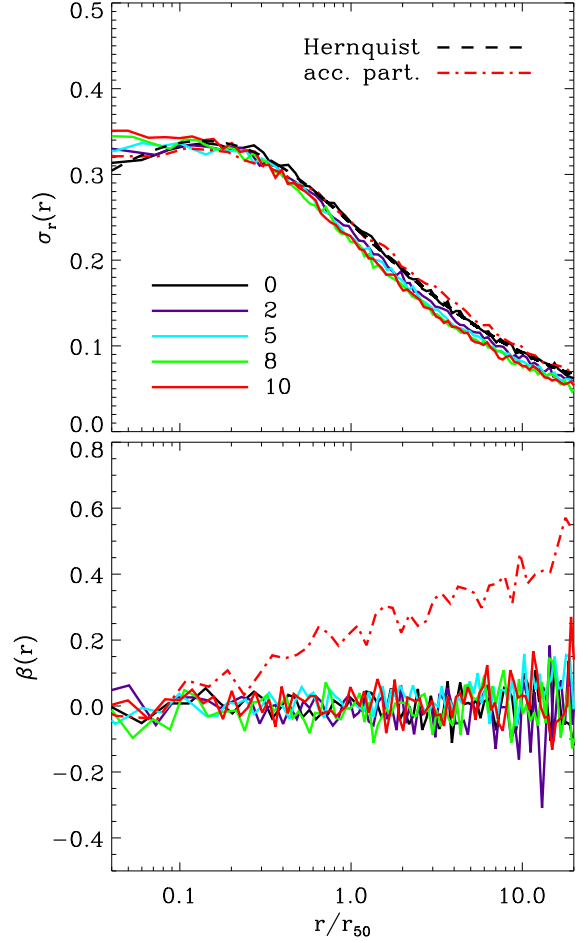}
  \end{center}
  \caption[title]
          {
            Top panel: The radial velocity dispersion for the head-on minor mergers
            of one-component models (B10hoc) stays constant over most of the radial range. 
            Only in the very central regions, it increases slightly with each generation.
            The black dashed line is the initial Hernquist profile and the red dashed-dotted
            line the velocity dispersion of all bound accreted particles.
            Bottom panel: For the whole bound remnant, the velocity 
            distribution stays perfectly isotropic, as the anisotropy parameter 
            $\beta$ stays zero. Looking at the accreted material (red dashed-dotted line)
            , it gets radially anisotropic with increasing radius. In both panels
            the radius is normalized to the spherical half-mass radius of the bound
            system.
          }
          \label{sig_acc}
\end{figure}

\begin{figure}
  \begin{center}
    \includegraphics[width=8cm]{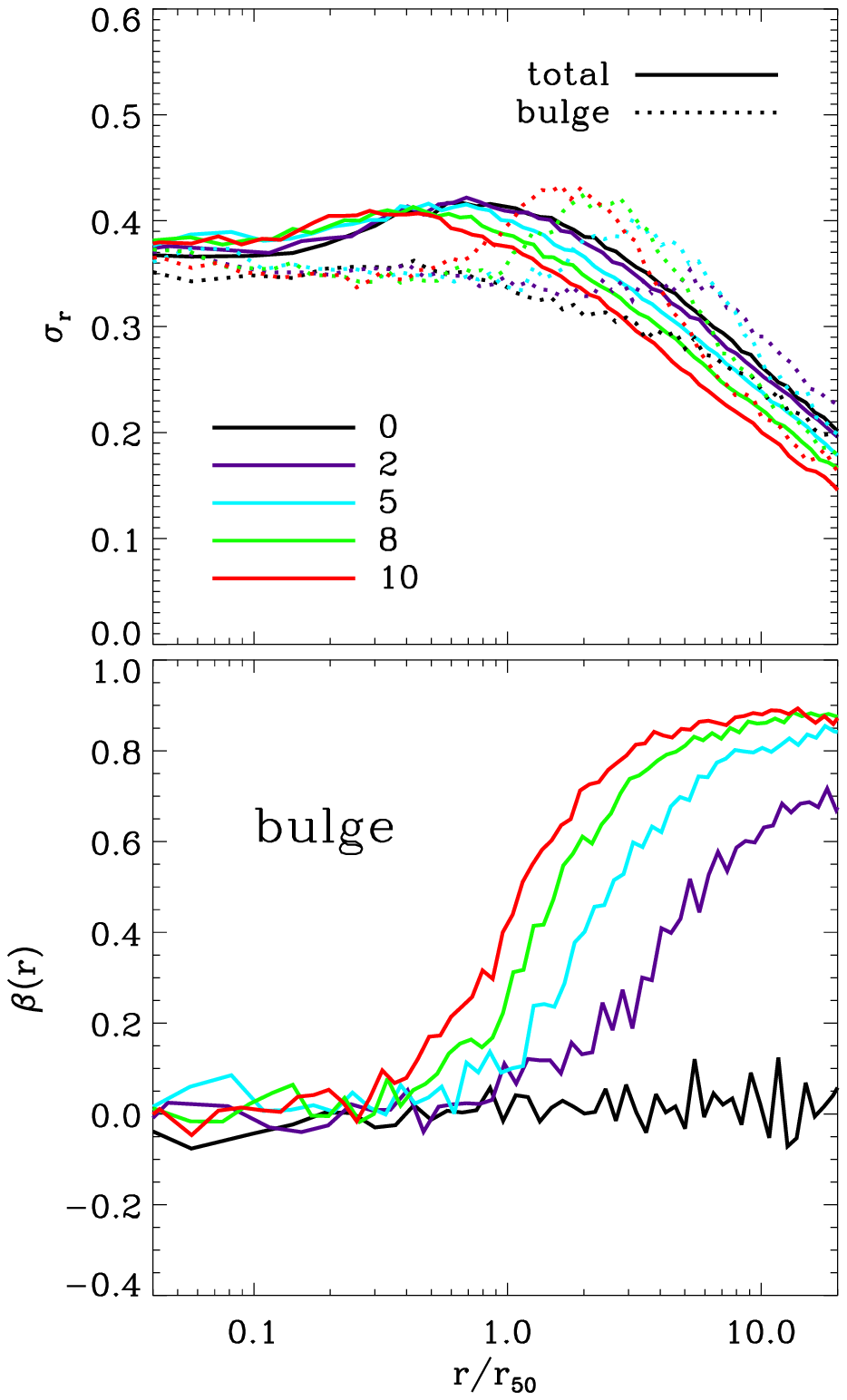}
  \end{center}
  \caption[title]
          {
           Top panel: The radial velocity dispersion of the total system (solid lines)
            for the head-on minor mergers of two-component models (HB10hod) stays constant 
            over the whole radial range. The dispersion of the bulge system (dotted line)
            builds up a prominent peak which comes from the accreted material, which 
            is mostly affected by dynamical friction. The radii are normalized to 
            the spherical half-mass radius of the bulge.
            Bottom panel: The anisotropy parameter of the bulge velocities gets radially
            biased at radii greater than the spherical half-mass radii of the bulge. 
          }
          \label{sig_acc2c}
\end{figure}

One prediction of violent relaxation is that during a merger the
system evolve to a state of higher entropy and the more complete the
violent relaxation is the higher the entropy, eventually approaching
the isothermal sphere. In reality, phase mixing damps the potential
fluctuations of a merger rapidly, and violent relaxation is
incomplete, which results in a  final equilibrium distribution which
does not reach the highest possible entropy state (see
\citet{1987MNRAS.229..103W} for a detailed discussion on maximum
entropy states in galaxies).   

In Fig. \ref{sigr1c} we show the radial velocity dispersion profiles
for major mergers of one-component systems. The radial velocity
dispersion of the final remnant (red line) can be well fitted with a
Jaffe profile \citep{1983MNRAS.202..995J} over a large radial
range. In the inner parts this profile resembles the singular
isothermal sphere \citep{1994AJ....107..634T}. The corresponding
changes of the structural parameter (see section \ref{VT}) are also 
given in the same figure. In the case of two-component models
(Fig. \ref{sigr1d}) we find the same result, i.e. the total (solid
line) and also the bulge profile (dotted line) approach a Jaffe
profile. \citet{1992ApJ...397L..75S} find the same trend for only one
generation of a head-on equal mass merger. This leads to the following
tentative conclusion. The evolution of the merger remnants towards a more
isothermal distribution is not unexpected since the entropy cannot
decrease in any collisionless process. However, by how much, or if at
all, the entropy increases is hard to anticipate since, formally,
there exists no maximum-entropy state for each merger generation. The
comparison with the Jaffe model shows that violent relaxation in the
centre is very effective. From one merger generation to the next both
the total mass and energy change. This provides new phase-space states
with ever higher binding energy that are populated by violent
relaxation. As a result, the nearly isothermal region of the remnant
extends deeper into the centre after each generation. 

In the bottom panels of Figs. \ref{sigr1c}, \ref{sigr1d} we can see 
that the initially isotropic remnants become radially anisotropic for
most of the galaxy $r > 0.1 r_{50}$. Already after the first merger
the anisotropy parameter  
\citep{2008gady.book.....B}
\begin{eqnarray}
  \beta=1-\frac{\sigma_{\theta}^2+\sigma_{\phi}^2}{2\sigma_{r}^2}. 
  \label{aniso}
\end{eqnarray}
becomes positive indicating radial anisotropy. This is not surprising,
as we only use orbits with very small or zero pericenter distances
(see also \citealt{2004MNRAS.349.1117B}). 

Next we focus on the evolution of the velocity dispersion profiles (Figs. \ref{sig_acc}, 
\ref{sig_acc2c}) of the minor mergers. From the previous analysis we
have seen that violent relaxation has no significant influence on the
central regions of the host galaxies.  
This is also reflected in the weak evolution of the central radial velocity
dispersion profiles  $\sigma_{r}(r)$ of the one- and two-component
minor mergers (top panels of Figs. \ref{sig_acc}, \ref{sig_acc2c}).  
In particular for the one-component scenario (e.g. B10hoc in top panel of Fig. 
\ref{sig_acc}) $\sigma_{r}(r)$ keeps the initial Hernquist profile
shape (black dashed line) over the whole radial range. Furthermore, as
$\beta(r) \sim 0$  (bottom panel Fig. \ref{sig_acc}) all remnants
(solid lines) stay isotropic. The accreted stars only show an
increasing radial anisotropy $\beta(r)>0$ with increasing
radius. The aforementioned effects are similar for all one-component
minor mergers, but slightly less pronounced in the cases with angular  
momentum. 

The merger remnants of two-component models (top panel Fig. \ref{sig_acc2c}) also 
indicate characteristics which are consistent with the evolution of
their differential energy distribution  (Fig. \ref{hist_ac2c}). The
central regions are hardly affected and therefore,  
the total velocity dispersion profile (solid lines in the top panel of
Fig. \ref{sig_acc2c}) and the one of the halo stay constant. The
accreted stellar particles create a bump in the energy distribution
(dashed lines in Fig. \ref{hist_ac2c}), which gets more and more prominent with each 
subsequent generation. These accreted particles induce an increasing
stellar velocity dispersion at radii larger than the spherical
half-mass radius $r_{50}$ (dotted lines, top panel 
Fig. \ref{sig_acc2c}). As the final coalescence of the stellar
component in the two-component scenario is on radial orbits,
independent of the initial conditions (see also  
\citet{2005MNRAS.361.1043G}), the anisotropy parameter becomes
radially biased for all of our minor mergers (e.g. HB10hod, bottom
panel of Fig. \ref{sig_acc2c}). This effect only occurs in the
simulation including a dark matter halo, because then the angular
momentum of the in-falling satellite is lost before the final merger,
due to enhanced dynamical friction. Hence, most of the stellar
particles approach on radial orbits and are, during the final
coalescence, stripped before reaching the center. If we use the
compact satellites, the overall trend does not change, but more
material reaches the center, as the particles are more tightly bound
and suffer less from tidal stripping.

\section{SUMMARY \& DISCUSSION}
\label{summary}

In this paper we have presented a series of numerical simulations of
major (mass-ratio of 1:1) and minor (mass-ratio of 1:10) mergers of
spherical, isotropic early-type galaxies. The model galaxies consist
of a bulge (one-component model) or a bulge embedded in a massive dark
matter halo (two-component model). After describing the procedure to
set up the initial conditions we have demonstrated the stability of
the one- and two-component galaxy models. By analysing the merger
simulations we have identified two different processes dominating the
evolution in sizes, velocity dispersion and dark matter fraction in 
major and minor mergers, respectively. 

Violent relaxation is the dominant mixing process in major mergers,
broadening the energy distribution, unbinding weakly bound particles
and making bound particles more bound. As a result the stellar velocity
dispersion increases and the size increase is weaker (r $\propto$
$M^{\alpha}$ with $\alpha <1$) than predicted from simple virial
arguments, an effect already reported by \citet{2003MNRAS.342..501N}. 
We present an analytical estimate for the size and dispersion
evolution taking the mass-loss into account and also separating the
effect on the dispersion of the stars and the dark matter halo for the
two-component models. The physical origin of the
increase of the dark matter fraction in equal-mass mergers reported by
\citet{2005MNRAS.362..184B} have been identified.  In the transition
region of luminous to dark matter domination, relatively more dark matter particles are
scattered from higher to lower energies (becoming more tightly
bound). This re-arrangement is driven by violent relaxation and leads
to an increase of the the dark matter fraction within the observable
effective radius of about 25 per cent for one equal-mass
merger, equivalent to a stellar mass increase of a factor of two.  

Stripping of stars from the low mass satellites is the dominant
process driving the size evolution in minor (1:10) mergers. The
stripped stars are predominantly deposited at large radii, leaving the
central structure of the host galaxies almost unchanged. This process
was  proposed to be the origin of abundance gradients in elliptical
galaxies \citep{1983MNRAS.204..219V}. The assembly of a 'halo' of
accreted stars leads to a rapid increase of the effective radius. In
systems with dark matter halos the stripped stars, with radially
biased velocity dispersion, are now populating regions where the host
galaxy was previously dominated by dark matter. As a
result the bulge size increase for two-component minor 
mergers is more rapid than predicted from simple analytical estimates (r
$\propto$ $M^{\alpha}$ with $\alpha >2$). This results in an
about three times  stronger increase of the dark matter fraction
within the stellar half-mass radius (compared to equal-mass
mergers) for a mass increases of a factor of two.   

Observational estimates of \citet{2010ApJ...709.1018V} indicate a size
growth of $r_{e}\;\alpha\;M^{2.04}$, which would be qualitatively
consistent with two-component minor merger scenarios. We have to note,
however, that the exact details will depend on the structure of the
galaxies, the merger orbits and the mass and extend of the dark matter
halos etc., a parameter space too large to cover in this
study. \citet{2009ApJ...706L..86N,2012arXiv1202.0971N}  followed a   
similar approach in a cosmologically motivated context and found a
much weaker size typical size increase ($r_{e}\propto M^{1.09}$). A
possible reason for this discrepancy is, that they averaged over all 
major and minor mergers. Additionally, they used a steeper central
slope for the stellar density profiles leading to more bound
satellites being even more compact than our satellites which lie on an
extrapolation of the $z=2$ mass-size relation of
\citet{2010ApJ...713..738W}. 

As we, and others before, have shown, increasing dark matter
fractions will change the ratio of dynamical to stellar mass. This
effect, if relevant in nature, will contribute to the observed tilt of
the fundamental plane, in addition to stellar populations and
structural non-homology
(e.g. \citealp{2003MNRAS.342..501N,2005MNRAS.362..184B,2007ApJ...658...65C,2010ApJ...717..803G,2011MNRAS.415..545T,2012arXiv1204.3099T}). We
also agree with \citet{2009ApJ...706L..86N} that the increase of the 
dark matter fraction is stronger for minor mergers and for this
scenario is dominated by the rapid size growth. But in contrast to
\citet{2009ApJ...706L..86N}, we find that the central dark matter
fraction of equal-mass mergers represents a 'real' change in the
internal structure caused by violent relaxation and not by the size
growth of the galaxies. More massive galaxies, which are expected to 
have acquired more stars in late mergers
\citep{2006MNRAS.366..499D,2006ApJ...648L..21K,2010ApJ...725.2312O,2012arXiv1205.5807M,2012arXiv1202.2357L},
will therefore have higher dark matter fractions (see also \citet{2010ApJ...712...88L}
for a detailed discussion).  This picture is in
agreement with recent high-resolution cosmological simulations 
(e.g. \citealp{2012arXiv1202.3441J}) and might explain the potentially
high dark matter fractions measured for elliptical galaxies in
particular if the merger history was dominated 
by minor mergers (e.g. \citealt{2010ApJ...712...88L,2011MNRAS.415.2215B,2010ApJ...724..511A}). 
This explanation for the tilt of the fundamental plane is
traditionally degenerate with uncertainties in the initial stellar mass
functions. Some recent studies imply that the stellar mass-to-light ratio
rises as a function of the stellar mass of the galaxies, leaving
relatively little room for the presence of dark matter at the center
of massive galaxies
\citep{2010Natur.468..940V,2012Natur.484..485C,2012arXiv1205.6473C,2012arXiv1205.6471V,2012arXiv1206.1594F}. 
If true this might be in considerable tension with cosmological
simulations of massive galaxies and a merger driven size growth
scenario - supported by massive dark matter halos - in general.  

The global mean square speeds of the systems decrease in minor mergers 
but the 'observed' effective line-of-sight velocity dispersions hardly
change. The most promising realistic scenario investigated here, the
two-component minor mergers, shows a relatively weak decrease in the
effective velocity dispersion, eventually too weak to explain the
available observational data
\citep{2009ApJ...696L..43C,2009Natur.460..717V,2011ApJ...736L...9V,2012arXiv1204.3099T}.   
However, reliable spectroscopic measurements are only available for few
galaxies so far and the constraints are weaker than for the size
evolution. In addition, studies like \citet{2009ApJ...696L..43C}
compared galaxies at a fixed stellar mass at different redshifts not
taking the expected mass increase into account, i.e. massive galaxies
at high redshift should be compared to even more massive galaxies
(with higher dispersions) at present. In addition, details of the
model, like mass concentration, orbits etc. will also affect the
dispersion evolution. It is, however, reassuring that more realistic
cosmological simulations, which also include a dissipative component,
are able to reproduce the observed dispersion evolution 
(e.g. \citealp{2012ApJ...744...63O}).

In summary our work shows that dissipationless dry minor mergers of
bulges embedded in dark matter halos can significantly increase the
sizes of a compact spheroids resembling high-redshift systems. As the
predictions for size growth are even above the observed values small
amount of gas, which is known to reduce the size growth \citep{2011MNRAS.415.3135C, 
2008ApJ...689...17H}, would perhaps not be enough to invalidate this
scenario. Recent simulations of galaxy formation in a full
cosmological context seem to support the conclusions drawn from the 
idealized simulations presented here with respect to the evolution of
sizes, dispersions, abundance gradient as well as dark matter fractions \citep{2007ApJ...658..710N,
2009ApJ...699L.178N,2009ApJ...692L...1J,2010ApJ...725.2312O,2012ApJ...744...63O,2010ApJ...709..218F,
2011ApJ...736...88F,2012arXiv1202.3441J,2012arXiv1206.0295L}.

\section*{Acknowledgments}

We thank Peter Johansson and Simon White for helpful discussions. This research was supported
by the DFG cluster of excellence Origin and Structure of the UniverseÓ and the DFG priority 
program SPP 1177.

\label{lastpage}

\end{document}